\documentclass[pre,showpacs,showkeys,preprintnumbers,amsmath,amssymb,superscriptaddress,twocolumn]{revtex4-2}
\usepackage{graphicx}
\usepackage{amsfonts}
\usepackage{url}

\usepackage{stackrel}
\usepackage{color}
\usepackage{epstopdf}
\usepackage{bm}

\usepackage[breaklinks=true,colorlinks=true,linkcolor=blue,urlcolor=blue,citecolor=blue]{hyperref}

\newcommand{\leftrarrows}{\mathrel{\raise.75ex\hbox{\oalign{%
  $\scriptstyle\leftarrow$\cr
  \vrule width0pt height.5ex$\hfil\scriptstyle\relbar$\cr}}}}
\newcommand{\lrightarrows}{\mathrel{\raise.75ex\hbox{\oalign{%
  $\scriptstyle\relbar$\hfil\cr
  $\scriptstyle\vrule width0pt height.5ex\smash\rightarrow$\cr}}}}
\newcommand{\Rrelbar}{\mathrel{\raise.75ex\hbox{\oalign{%
  $\scriptstyle\relbar$\cr
  \vrule width0pt height.5ex$\scriptstyle\relbar$}}}}
\newcommand{\longleftrightarrows}{\leftrarrows\joinrel\Rrelbar\joinrel\lrightarrows}

\makeatletter
\def\leftrightarrowsfill@{\arrowfill@\leftrarrows\Rrelbar\lrightarrows}
\newcommand{\xleftrightarrows}[2][]{\ext@arrow 3399\leftrightarrowsfill@{#1}{#2}}
\makeatother

\def\x{\bm{x}}
\def\X{\bm{X}}

\def\E{\mathbb{E}}
\def\P{\mathbb{P}}
\def\R{\mathbb{R}}
\def\K{\mathcal{K}}
\def\L{\mathcal{L}}
\def\M{\mathcal{M}}
\def\N{\mathcal{N}}
\def\T{\mathcal{T}}
\def\pa{\partial\Omega}

\def\erf{\mathrm{erf}}
\def\erfcx{\mathrm{erfcx}}

\def\koff{k_{\rm off}}
\def\kon{k_{\rm on}}

\def\ddelta{t_d}
\def\ttau{t_a}

\newcommand{\dtilde}[1]{\bar{#1}}

\begin{document}

\title{Diffusion-controlled reactions with \\ non-Markovian binding/unbinding kinetics}

\author{Denis~S.~Grebenkov}
 \email{denis.grebenkov@polytechnique.edu}
\affiliation{
Laboratoire de Physique de la Mati\`{e}re Condens\'{e}e, \\ 
CNRS -- Ecole Polytechnique, Institut Polytechnique de Paris, 91120 Palaiseau, France}

\date{\today}

\begin{abstract}
We develop a theory of reversible diffusion-controlled reactions with
generalized binding/unbinding kinetics.  In this framework, a
diffusing particle can bind to the reactive substrate after a random
number of arrivals onto it, with a given threshold distribution.  The
particle remains bound to the substrate for a random waiting time
drawn from another given distribution and then resumes its bulk
diffusion until the next binding, and so on.  When both distributions
are exponential, one retrieves the conventional first-order forward
and backward reactions whose reversible kinetics is described by
generalized Collins-Kimball's (or back-reaction) boundary condition.
In turn, if either of distributions is not exponential, one deals with
generalized (non-Markovian) binding or unbinding kinetics (or both).
Combining renewal technique with the encounter-based approach, we
derive spectral expansions for the propagator, the concentration of
particles, and the diffusive flux on the substrate.  We study their
long-time behavior and reveal how anomalous rarity of binding or
unbinding events due to heavy tails of the threshold and waiting time
distributions may affect such reversible diffusion-controlled
reactions.  Distinctions between time-dependent reactivity,
encounter-dependent reactivity, and a convolution-type Robin boundary
condition with a memory kernel are elucidated.
\end{abstract}





\maketitle

\section{Introduction}
\label{sec:intro}

The importance of diffusion in chemical reactions was first recognized
by von Smoluchowski \cite{Smoluchowski17} and then became a broad
field of intensive research in physical chemistry
\cite{Rice,House,Metzler,Lindenberg}.  In many chemical and
biophysical processes, the involved species (atoms, ions, molecules,
or even whole organisms such as bacteria) have first to encounter each
other, or to find a specific target or a substrate, to initiate a
reaction event \cite{Alberts,Lauffenburger,Bressloff13}.  Most former
works were dedicated to understand the role of this first-passage step
in chemical reactions \cite{Redner,Schuss}.  In particular, the
dependence of the first-passage time (FPT) distribution on the
geometric structure of the environment
\cite{Benichou10,Benichou14,Vaccario15,Grebenkov16,Godec16b,Galanti16},
on the size and shape of the target region
\cite{Kolokolnikov05,Singer06a,Singer06b,Singer06c,Schuss07,Benichou08,Pillay10,Cheviakov10,Cheviakov11,Cheviakov12,Holcman13,Holcman14,Isaacson16,Grebenkov17a,Grebenkov17b,Agranov18},
and on the type of diffusion process \cite{Condamin07,Lanoiselee18}
was thoroughly investigated.  For instance, the impact of anomalously
long halts or non-Markovian bulk dynamics onto diffusion-controlled
reactions (also known as diffusion-influenced, diffusion-limited or
diffusion-mediated reactions) was emphasized
\cite{Grebenkov10a,Grebenkov10b,Guerin16}.  Eventual limitations of
the mean FPT and the importance of knowing whole distribution of FPT
for an adequate description of chemical reactions were discussed
\cite{Mattos12,Godec16,Grebenkov18,Grebenkov18b,Grebenkov19c}.

While the importance of diffusion in the bulk is now fully
acknowledged, the role of surface reactions, occurring after the
arrival of a molecule onto the reactive substrate (or after the
encounter of two molecules), still remains underestimated.  Collins
and Kimball realized already in 1949 that the original assumption by
Smoluchowski of a perfect reaction upon the first encounter of two
species was too idealized and limited in practice \cite{Collins49}.  In
fact, the molecules may need to overcome an activation barrier or to
be in appropriate conformational states to be able to react.  When the
reaction is understood as as escape from a confining domain through a
hole, an entropic barrier has to be overcome
\cite{Zhou91,Reguera06,Malgaretti16,Skvortsov23}.  Moreover, if the
escape region is not simply a ``hole'' but, e.g., an ion channel, it
has to be in an ``open'' (or active) state when the ion arrives
\cite{Benichou00,Reingruber09,Lawley15,Bressloff17}.  In
heterogeneous catalysis, the particle may arrive onto the catalytic
surface at a passive (or passivated) location and thus resume its
diffusive search for an active site
\cite{Berg77,Shoup81,Zwanzig91,Berezhkovskii06,Muratov08,Filoche08,Bernoff18,Punia21,Chaudhury23}.
Whatever the microscopic mechanism of surface reaction is, the arrived
particle may either react or be reflected back to resume its bulk
diffusion.  As a consequence, the successful reaction event is
generally preceded by a long sequence of failed reaction attempts and
the consequent diffusive explorations of the bulk.  This stochastic
process is known as partially reflected Brownian motion
\cite{Grebenkov06,Grebenkov07} or partially reflected diffusion
\cite{Singer08}.  

Collins and Kimball suggested to employ the Robin (also known as
radiation) boundary condition to describe the concentration of
diffusing particles, $c(\x,t)$, on the partially reactive region
$\pa_R$:
\begin{equation}  \label{eq:Robin}
-D \partial_n c(\x,t) = \kappa_0 \, c(\x,t)  \quad (\x\in\pa_R),
\end{equation}
where $D$ is the (self-)diffusion coefficient of particles,
$\partial_n$ is the normal derivative on the boundary oriented
outwards the confining domain, and $\kappa_0$ is the reactivity of the
region.  This condition {\it postulates} that the net diffusive flux
of particles from the bulk onto the reactive region (the left-hand
side) is proportional to the concentration on that surface (the
right-hand side).  The proportionality coefficient $\kappa_0$ can
range from $0$ for the no reaction setting on an inert impermeable
surface, to $+\infty$ for an immediate reaction upon the first
encounter with a perfectly reactive region, as formulated by
Smoluchowski.
Such an irreversible process can be schematically described by either
of two reactions
\begin{equation}   \label{eq:reaction_irrev}
A + C \stackrel[\kon]{}\longrightarrow [AC]  \quad \textrm{or}  \quad
A + C \stackrel[\kon]{}\longrightarrow B + C,
\end{equation}
where $A$ denotes the diffusing molecule, $C$ is the immobile reactive
region, $[AC]$ is the formed complex (if the molecule $A$ stuck on
$C$), and $B$ is the product of chemical transformation of $A$ on $C$,
or the same particle that has lost its excited state (e.g., a
transverse magnetization of a spin-bearing particle can be lost due to
surface relaxation or an excited luminescent state of a nanoparticle
can relax).  The role of reactivity onto diffusion-controlled
reactions was thoroughly investigated
\cite{Sano79,Brownstein79,Weiss86,Powles92,Sapoval94,Sapoval02,Grebenkov05,Traytak07,Bressloff08,Grebenkov15,Serov16,Piazza19,Guerin21}.
In particular, the bimolecular rate constant $\kon$ (in units
m$^3$/mol/s or 1/M/s) is proportional to the reactivity $\kappa_0$ (in
units m/s):
\begin{equation}
\kon = \kappa_0 N_A |\pa_R| ,
\end{equation}
where $N_A$ is the Avogadro number, and $|\pa_R|$ is the surface area
of the reactive region.  Bearing in mind this relation, we will use
the reactivity $\kappa_0$ in the following discussions.

The next step consists in accounting for reversible binding.  In fact,
the Robin boundary condition describes the reactive region as a sink
or a definitive trap for the reacted particle, as if it was killed,
destroyed, irreversibly transformed into another molecule, or
irreversibly lost its excited state.  In many applications, however,
the chemical reaction can be reversed, while the particle that was
adsorbed onto a substrate, can desorb from it and resume its bulk
diffusion.  In other words, the forward reactions in
Eq. (\ref{eq:reaction_irrev}) have to be completed by backward
reactions, which are typically characterized by the backward reaction
rate (also known as desorption or dissociation rate) $\koff$:
\begin{equation}  \label{eq:reaction_rev}
A + C \stackrel[\kon]{\koff}\longleftrightarrows [AC]   \quad \textrm{or}  \quad
A + C \stackrel[\kon]{\koff}\longleftrightarrows B + C .
\end{equation}
In a broad sense, such bimolecular reversible reactions can also be
understood as an adsorption/desorption process of a molecule $A$ on a
substrate $C$.  While the microscopic physochemical mechanisms of
these processes are different, their effect onto the concentration
admits the same mathematical description.  Note also that
adsorption/desorption processes were usually considered on a flat
surface and thus modeled by one-dimensional diffusion, whereas
bimolecular reactions were most often associated to a spherical
geometry.  In general, however, an adsorbing surface of a porous
medium can be non-flat even at nanoscopic scales, while many proteins
and other macromolecules are not spherical.  In this light, one needs
to go beyond the conventional settings and to treat
diffusion-controlled reactions in general domains.  Despite
microscopic differences between reversible bimolecular reactions and
adsorption/desorption processes, we will use inter-changeably the
terms ``forward reaction'', ``binding'', ``adsorption'',
``association'' and ``recombination'', as well as the terms ``backward
reaction'', ``unbinding'', ``desorption'', and ``dissociation''.

A theoretical description of reversible reactions and
adsorption/desorption processes in Eq. (\ref{eq:reaction_rev}) is well
established (see
\cite{Ward46,Goodrich54,Baret68,Mysels82,Frisch83,Borwankar83,Agmon84,Adamczyk87,Agmon88,Agmon89,Agmon90,Agmon93,Chang95,Liggieri96,Kim99,Foo10,Prustel13,Prustel13b,Miura15,Grebenkov19e,Scher23}
and references therein).  In a nutshell, one introduces the surface
concentration of particles in the bound state, $c_b(\x,t)$, and writes
the first-order exchange kinetics at each boundary point on the
reactive region:
\begin{equation}  \label{eq:Robin_rev1}
\partial_t c_b(\x,t) = \kappa_0\, c(\x,t) - \koff \, c_b(\x,t)   \quad (\x\in\pa_R), 
\end{equation}
which describes both forward reaction (the first term) and backward
reaction (the second term).  In adsorption theory, the linear form of
the forward term is known as the Henry's law \cite{Chang95} (see also
an overview in \cite{Scher23}).
At the same time, the conservation of mass implies that any change in
the fraction of bound particles is equal to the diffusive flux density
of unbound particles from the bulk to the reactive region:
\begin{equation}   \label{eq:Robin_rev2}
\partial_t c_b(\x,t) = - D\partial_n c(\x,t)  \quad (\x\in\pa_R).
\end{equation}
Equating the right-hand sides of these equations, one gets the
``back-reaction'' boundary condition, also known as ``generalized
radiation'' or ``generalized Collins-Kimball'' boundary condition
\cite{Goodrich54,Agmon84,Kim99,Prustel13,Grebenkov19e}.  Microscopic
derivation and interpretation of Eqs. (\ref{eq:Robin_rev1},
\ref{eq:Robin_rev2}) are discussed in Appendix \ref{sec:micro}.

These two equations can be reduced to a single Robin-type boundary
condition by taking the Laplace transform (denoted by tilde) with
respect to time $t$, e.g.,
\begin{equation}
\tilde{c}(\x,p) = \L\{ c(\x,t)\} = \int\limits_0^\infty dt \, e^{-pt} \, c(\x,t).
\end{equation}
Rewriting Eqs. (\ref{eq:Robin_rev1}, \ref{eq:Robin_rev2}) in the
Laplace domain and eliminating $\tilde{c}_b(\x,p)$ from them yield
\begin{equation}  \label{eq:Robin_rev}
-D \partial_n \tilde{c}(\x,p) = \tilde{\kappa}(p) \, \tilde{c}(\x,p) \quad (\x\in\pa_R),
\end{equation}
with
\begin{equation}
\tilde{\kappa}(p) = \frac{\kappa_0}{1 + \koff/p} \,,
\end{equation}
where we assumed that there was no bound particle in the initial state
(i.e., $c_b(\x,0) = 0$).  In the Laplace domain, the effect of
reversible binding is thus incorporated through the $p$-dependent
reactivity $\tilde{\kappa}(p)$.  In other words, if one knows the
solution for irreversible binding, it is enough to replace $\kappa_0$ by
$\tilde{\kappa}(p)$ to incorporate backward reactions.  

In turn, irreversible and reversible settings differ in time domain;
in fact, the inverse Laplace transform of Eq. (\ref{eq:Robin_rev})
yields a convolution-type boundary condition
\begin{equation}  \label{eq:Robin_conv}
- D \partial_n c(\x,t) = \int\limits_0^t dt' \, \K(t-t') \, c(\x,t')  \quad (\x\in\pa_R),
\end{equation}
with a memory kernel 
\begin{equation}  \label{eq:kappa_exp}
\K(t) = \L^{-1}\{ \tilde{\kappa}(p)\} = \kappa_0 \bigl( \delta(t) - \koff \, e^{-\koff t}\bigr),
\end{equation}
where $\L^{-1}$ denotes the inverse Laplace transform.  In contrast to
Eq. (\ref{eq:Robin}) for irreversible binding, the boundary condition
(\ref{eq:Robin_conv}) is nonlocal in time: a particle that was
adsorbed at an earlier time $t'$ can desorb at a later time $t$ and
thus contribute to the net diffusive flux in the left-hand side.  The
effect of binding/unbinding events onto the time evolution of the
concentration $c(\x,t)$ and on other first-passage events was
investigated.  Moreover, simultaneous binding of multiple
independently diffusing particles is known to control many activation
mechanisms in microbiology such as signaling in neurons, synaptic
plasticity, cell apoptosis caused by double strand DNA breaks, cell
differentiation and division
\cite{Berridge03,Duc10,Guerrier16,Reva21}.  The random, asynchronous
binding/unbinding events for each particle bring new statistical
challenges to the theoretical description of such systems
\cite{Grebenkov17e,Lawley19,Grebenkov22e,Grebenkov22f}.

In summary, most theoretical works on reversible reactions were
limited to the above setting, in which binding events are incorporated
via Robin boundary condition with a single forward reaction constant
$\kon$ (or $\kappa_0$), while unbinding events are characterized by a
first-order kinetics with a single rate $\koff$.
Despite the convolution-type boundary condition (\ref{eq:Robin_conv})
for the concentration $c(\x,t)$ alone, this kinetics is called
Markovian because the time evolution of the particle state,
characterized by the pair $c(\x,t)$ and $c_b(\x,t)$, is governed by
the diffusion equation $\partial_t c(\x,t) = D\Delta c(\x,t)$ and the
differential equation (\ref{eq:Robin_rev1}) that are both local in
time; as a consequence, the future state depends on the present but
not on the past.

In this paper, we aim at extending the current framework to much more
general non-Markovian surface reactions.  On the one hand, each bound
state can be characterized by a random waiting time drawn from a
prescribed probability density $\phi(t)$.  The above first-order
kinetics of backward reactions would then correspond to an exponential
distribution, with
\begin{equation}  \label{eq:phi_exp}
\phi(t) = \koff \, e^{-\koff t}.  
\end{equation}
This extension was introduced by Agmon and Weiss for a pair of
reversibly binding particles whose motion is described by a
spherically symmetric diffusion equation \cite{Agmon89}.  We will
further extend this concept to general confining domains.

On the other hand, one can also modify the description of the forward
reaction by using the encounter-based approach \cite{Grebenkov20}.
This approach relies on the notion of the boundary local time
$\ell_t$, i.e., a rescaled number of encounters between the diffusing
particle and the reactive region (see Sec. \ref{sec:encounter} for
details).  As the particle attempts to bind at each encounter, the
successful binding occurs when the number of failed attempts,
characterized by $\ell_t$, exceeds some random threshold $\hat{\ell}$
described by a probability density $\psi(\ell)$.  If the random
threshold obeys the exponential probability density
\begin{equation}  \label{eq:psi_exp}
\psi(\ell) = q\, e^{-q\ell}  \quad (q = \kappa_0/D), 
\end{equation}
one retrieves the standard description of binding events via Robin
boundary condition (\ref{eq:Robin}), as detailed below in
Sec. \ref{sec:Robin}.  In analogy to Eq. (\ref{eq:phi_exp}), we call
the binding kinetics characterized by Eq. (\ref{eq:psi_exp}) as
Markovian.  In turn, other probability densities $\psi(\ell)$ yield
non-Markovian binding kinetics.  One of the crucial advantages of the
encounter-based approach is that both binding and unbinding events can
be characterized in a very similar way.

The paper is organized as follows.  Section \ref{sec:theory} presents
the general theory.  After a brief introduction of the main
ingredients of the encounter-based approach in
Sec. \ref{sec:encounter}, we employ the renewal technique to derive
the propagator in Sec. \ref{sec:main}.  Section \ref{sec:bindingM} is
dedicated to the simpler case when only unbinding events are
non-Markovian, while Sec. \ref{sec:Robin} illustrates the violation of
the Robin boundary condition in the general case of non-Markovian
binding.  Different long-time regimes are discussed in
Sec. \ref{sec:long}.  Section \ref{sec:sphere} illustrates the general
properties for the case of a spherical reactive region.  In
particular, we inspect the probability to be in the unbound state, the
diffusive flux and the concentration profile.  In
Sec. \ref{sec:discussion}, we discuss limitations of the conventional
description and its extensions via different types of variable
reactivity.  Section \ref{sec:conclusion} concludes the paper, with
the summary of main results and further perspectives.

\section{Theory}
\label{sec:theory}

\begin{figure}[t!]
\begin{center}
\includegraphics[width=65mm]{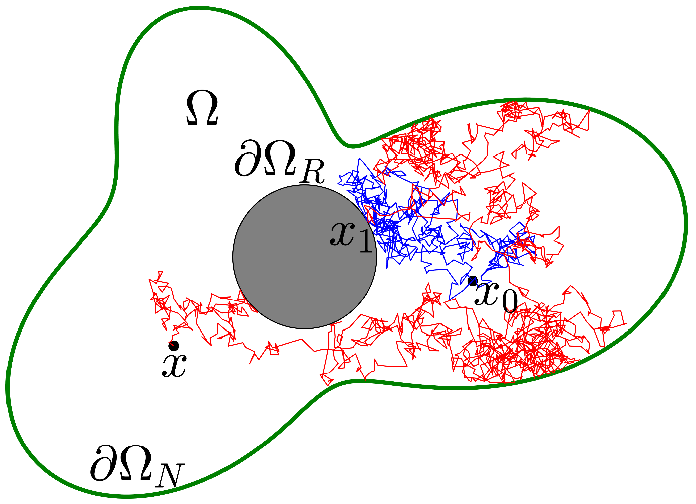} 
\end{center}
\caption{
Illustration of a confining domain $\Omega$ with a smooth boundary
$\pa$, which is split into two disjoint parts: a reflecting boundary
$\pa_N$ (in green) and a reactive region $\pa_R$ (the frontier of a
gray obstacle).  A simulated trajectory, started from $\x_0$ and
arrived in $\x$, exhibits multiple failed attempts to react on $\pa_R$
before the successful reaction (binding) at an intermediate point
$\x_1$ (path in blue).  After staying in the bound state for a random
waiting time, the particle was released from $\x_1$ and resumed its
bulk diffusion toward the arrival point $\x$ (path in red).  }
\label{fig:domain}
\end{figure}

We consider ordinary diffusion of a point-like particle with a
constant diffusivity $D$ inside a bounded Euclidean domain $\Omega
\subset\R^d$ with a smooth boundary $\pa$, which is split into two
disjoint parts: a passive (inert) region $\pa_N$, which confines the
particle by reflecting it back to the domain, and a reactive region
$\pa_R$, to which the particle can bind reversibly
(Fig. \ref{fig:domain}).  The binding mechanism is characterized by a
given probability density $\psi(\ell)$ of a random threshold
$\hat{\ell}$, as described explicitly in Sec. \ref{sec:encounter}.
After each binding event, the particle remains immobile on the
reactive region for a random waiting time $\hat{t}$ drawn from a given
probability density $\phi(t)$.  When this time elapses, the particle
unbinds and resumes its bulk diffusion from the boundary point where
it was bound.  We assume that all binding and unbinding events occur
independently from each other.

At a single-molecule level, we aim at obtaining the propagator
$Q(\x,t|\x_0)$, i.e., the probability density that a single particle
started from $\x_0 \in\Omega$ at time $0$ is found in the unbound
state in a vicinity of point $\x\in \Omega$ at time $t$.  We will also
discuss the probability of finding the particle in the unbound state
at time $t$:
\begin{equation}
S(t|\x_0) = \int\limits_{\Omega} d\x \, Q(\x,t|\x_0),
\end{equation}
whereas $1-S(t|\x_0)$ is the probability to be in the bound state at
time $t$.  For irreversible reactions, $S(t|\x_0)$ is called the
survival probability.

At a macroscopic level, if there are many independently diffusing
particles with a given initial concentration $c_0(\x_0)$, the
propagator determines the concentration of these particles at time $t$
as
\begin{equation}  \label{eq:cx_Q}
c(\x,t) = \int\limits_{\Omega} d\x_0 \, Q(\x,t|\x_0) \, c_0(\x_0).
\end{equation}
In turn, 
\begin{equation}
J(t) = \int\limits_{\pa_R} d\x \, (- D\partial_n c(\x,t))
\end{equation}
is the diffusive flux of particles onto the reactive region.  We will
derive all these quantities and discuss their behavior.

\subsection{Binding events}
\label{sec:encounter}

The encounter-based approach employs the statistics of encounters
between the diffusing particle and the reactive region $\pa_R$ to
implement surface reactions \cite{Grebenkov20}.  As the details of
this approach were presented in earlier publications
\cite{Grebenkov20,Grebenkov20b,Grebenkov20c,Bressloff22a,Bressloff22c},
we just sketch the main steps and ``ingredients'' needed for our
computation.

In addition to the (random) position $\X_t$ of the particle at time
$t$, we introduce its boundary local time $\ell_t$ on $\pa_R$ as
\begin{equation}
\ell_t = \lim\limits_{a\to 0} a\, \N_t^{(a)} ,
\end{equation}
where $\N_t^{(a)}$ is the (random) number of downcrossings of a thin
layer of width $a$ near $\pa_R$ up to time $t$.  Each such
downcrossing can be understood as an encounter between the particle
and the substrate $\pa_R$ at a given spatial resolution $a$.  As $a\to
0$, $\N_t^{(a)}$ diverges due to the self-similar character of
Brownian motion but its rescaling by $a$ yields a nontrivial limit
$\ell_t$, which is a non-decreasing process (with $\ell_0 = 0$) that
increments at each encounter of the particle with $\pa_R$
\cite{Levy,Ito,Freidlin}.
The boundary local time should not be confused with the local time in
a bulk point, which was thoroughly studied (see
\cite{Borodin,Majumdar05} and references therein).  Despite its name,
$\ell_t$ has units of length.  While physical time $t$ is a proxy of
the number of particle's jumps in the bulk, the boundary local time
$\ell_t$ characterizes the number of jumps onto the surface
\cite{Grebenkov20b}.  In this light, both quantities are complementary
and tightly related (see \cite{Grebenkov20} for further discussion).
The diffusive dynamics of the particle is fully described by the pair
$(\X_t,\ell_t)$, whose probability density is called the full
propagator $P(\x,\ell,t|\x_0)$.  The crucial distinction of the
encounter-based approach from conventional methods is that the full
propagator characterizes purely diffusive dynamics by treating the
region $\pa_R$ as reflecting.  In turn, surface reactions on $\pa_R$
are then introduced explicitly via an appropriate stopping condition.

In fact, as discussed earlier, the reaction event is preceded by a
sequence of failed reaction attempts at each encounter with $\pa_R$.
In other words, the reaction occurs at the first instance $\T$ when
the boundary local time (the proxy of the failed attempts) exceeds
some threshold $\hat{\ell}$:
\begin{equation}  \label{eq:Tdef}
\T = \inf\{ t > 0 ~:~ \ell_t > \hat{\ell}\} .
\end{equation}
The choice of this threshold selects the mechanism of surface reaction
\cite{Grebenkov20}.  For instance, if $\hat{\ell} = 0$, the reaction
occurs when the boundary local time $\ell_t$ first exceeds $0$, i.e.,
when the particle first encounters the region $\pa_R$.  This is the
perfect reaction introduced by Smoluchowski \cite{Smoluchowski17}.  In
turn, if the particle attempts to react independently on each
encounter with probability $\rho \approx a\kappa_0/D \ll 1$, the
probability of no surface reaction up to the $n$-th encounter is
\begin{equation}
1 - \sum\limits_{k=1}^n \rho (1-\rho)^{k-1} = (1-\rho)^n \approx e^{-\rho n} \approx e^{-q \ell}, 
\end{equation}
where $q = \kappa_0/D$, $\ell = na$, and the auxiliary parameter $a$
is a small width of the reactive layer (which is needed here as a sort
of regularization but then eliminated in the limit $a\to 0$).  By
introducing an exponentially distributed random threshold $\hat{\ell}$
such that $\P\{\ell < \hat{\ell}\} = e^{-q\ell}$ and associating the
$n$-th reaction attempt to the number of encounters $\N_t^{(a)}
\approx \ell_t/a$ up to time $t$, one sees that the condition $\ell_t
< \hat{\ell}$ incorporates the survival of the particle in the
presence of surface reactions on $\pa_R$ with a constant reactivity
$\kappa_0$.  Most importantly, the encounter-based approach allows one
to go far beyond this classical mechanism and to describe much more
general and sophisticated surface reactions by means of a random
threshold $\hat{\ell}$ obeying a general probability density
$\psi(\ell)$, as discussed in this paper.

Let $G_\psi(\x,t|\x_0)$ denote the generalized propagator in the
presence of {\it irreversible} forward reaction on $\pa_R$, i.e.,
$G_\psi(\x,t|\x_0)$ is the probability density of diffusion from
$\x_0$ to $\x$ in time $t$ under the survival condition (no forward
reaction on $\pa_R$ up to time $t$).  As the survival condition can be
written in terms of the boundary local time, $\ell_t < \hat{\ell}$,
one gets
\begin{equation}
G_\psi(\x,t|\x_0) = \int\limits_0^\infty d\ell \, \Psi(\ell) \, P(\x,\ell,t|\x_0),
\end{equation}
where 
\begin{equation}
\Psi(\ell) = \P\{\ell < \hat{\ell}\} = \int\limits_{\ell}^\infty d\ell' \, \psi(\ell').
\end{equation}
In other words, the full propagator $P(\x,\ell,t|\x_0)$ that treats
$\pa_R$ as reflecting, is explicitly complemented by the survival
condition $\ell_t < \hat{\ell}$, which is fulfilled with probability
$\Psi(\ell)$ for a given realized value $\ell$ of the boundary local
time $\ell_t$.  The above integral sums up contributions from all
possible realizations of $\ell_t$.  The associated probability flux
density on the reactive region $\pa_R$ reads
\begin{equation}  
j_\psi(\x,t|\x_0) = -D\partial_n G_\psi(\x,t|\x_0)  \quad (\x\in \pa_R).
\end{equation}
This quantity characterizes a single binding event (the unbinding
mechanism will be introduced in Sec. \ref{sec:main}).

In the case of Markovian binding, Eq. (\ref{eq:psi_exp}) implies
$\Psi(\ell) = e^{-q\ell}$, so that one retrieves the conventional
propagator 
\begin{equation}
G_q(\x,t|\x_0) = \int\limits_0^\infty d\ell \, e^{-q\ell} \, P(\x,\ell,t|\x_0),
\end{equation}
that satisfies the Robin boundary condition on $\pa_R$ (see
\cite{Grebenkov20})
\begin{equation}  \label{eq:Gq_Robin}
-\partial_n G_q(\x,t|\x_0) = q \, G_q(\x,t|\x_0)  \quad (\x\in \pa_R),
\end{equation}
and the corresponding probability flux density 
\begin{equation}
j_q(\x,t|\x_0) = -D\partial_n G_q(\x,t|\x_0) \quad (\x\in\pa_R). 
\end{equation}
We will often refer to two particular limits: $q = \infty$ and $q =
0$.  The quantities $G_q(\x,t|\x_0)$ and $j_q(\x,t|\x_0)$ can be
represented via spectral expansions over the eigenvalues and
eigenfunctions of the Laplace operator
\cite{Redner,Gardiner,VanKampen}.  Unfortunately, these commonly used
eigenfunctions are not suitable to access the generalized propagator
$G_\psi(\x,t|\x_0)$ that does not satisfy the Robin boundary condition
(\ref{eq:Gq_Robin}). 

This problem was solved in \cite{Grebenkov20} by using the
Dirichlet-to-Neumann operator $\M_p$.  This operator acts on an
appropriate space of functions on $\pa_R$ as follows: for a function
$f$ on $\pa_R$, one finds the solution $u$ of the boundary value
problem
\begin{subequations}
\begin{align}
(p - D \Delta) u & = 0 \quad (\x\in\Omega), \\  \label{eq:u_D}
u & = f \quad (\x\in\pa_R) , \\   \label{eq:u_N}
\partial_n u & = 0 \quad (\x\in\pa_N),
\end{align}
\end{subequations}
evaluates its normal derivative on $\pa_R$, $(\partial_n u)|_{\pa_R}$,
and associates it to $f$: 
\begin{equation}
\M_p ~:~ f \to (\partial_n u)|_{\pa_R} \quad \textrm{or} \quad
\M_p f = (\partial_n u)|_{\pa_R}.
\end{equation}
Intuitively, the operator $\M_p$ transforms the Dirichlet boundary
condition $u = f$ on $\pa_R$ to an equivalent Neumann boundary
condition $\partial_n u = \M_p f$ on $\pa_R$.  In physical terms, if
$f$ is understood as a density of particles maintained on $\pa_R$, the
operator $\M_p$ determines their steady-state diffusive flux density
$D \M_p f = D(\partial_n u)|_{\pa_R}$ into the reactive bulk (with a
bulk reaction rate $p$).  In the particular case $p = 0$, a similar
problem arises in heat transfer from the boundary $\pa_R$ with an
imposed temperature profile $f$, so that $\M_0 f$ is proportional to
the steady-state temperature flux density into the bulk; moreover, the
analogy with electrostatics allows one to interpret $f$ as a charge
density, for which $\M_0 f$ determines the current density.

The Dirichlet-to-Neumann operator was thoroughly investigated for the
case when $\pa_R = \pa$ (i.e., $\pa_N = \emptyset$)
\cite{Arendt14,Daners14,terElst14,Behrndt15,Arendt15,Hassell17,Girouard17}.
We conjecture that the inclusion of the additional boundary condition
(\ref{eq:u_N}) on the reflecting part $\pa_N$ does not affect general
properties of this operator; in particular, as the region $\pa_R$ is
bounded, $\M_p$ remains to be self-adjoint pseudo-differential
operator with a discrete spectrum for any $p \geq 0$.  Its eigenvalues
can be enumerated in an ascending order,
\begin{equation}
0 \leq \mu_0^{(p)} \leq \mu_1^{(p)} \leq \ldots \leq \mu_k^{(p)} \leq \ldots \nearrow +\infty ,
\end{equation}
while the eigenfunctions $\{v_k^{(p)}\}$ form a complete orthonormal
basis in the space $L_2(\pa_R)$ of square integrable functions on
$\pa_R$.  While we checked these spectral proprieties numerically for
various domains (see an example in Sec. \ref{sec:sphere}), a
mathematical proof of this extension is beyond the scope of this
paper.

The following spectral expansion of the generalized propagator in the
Laplace domain was derived in \cite{Grebenkov20}: 
\begin{align}  \label{eq:Gpsi}
\tilde{G}_\psi(\x,p|\x_0) & = \tilde{G}_\infty(\x,p|\x_0) \\  \nonumber
& + \frac{1}{D} \sum\limits_{k=0}^\infty [V_k^{(p)}(\x_0)]^* V_k^{(p)}(\x) \frac{1 - \dtilde{\psi}(\mu_k^{(p)})}{\mu_k^{(p)}} \,,
\end{align}
where asterisk denotes complex conjugate, 
\begin{equation} 
\dtilde{\psi}(\mu) = \int\limits_0^\infty d\ell \, e^{-\mu \ell} \, \psi(\ell)
\end{equation}
is the Laplace transform of $\psi(\ell)$ (we use bar instead of tilde
here to highlight different units),
\begin{equation}
V_k^{(p)}(\x_0) = \int\limits_{\pa_R} d\x \, \tilde{j}_\infty(\x,p|\x_0) \, v_k^{(p)}(\x),
\end{equation}
and $\tilde{G}_\infty(\x,p|\x_0)$ and $\tilde{j}_\infty(\x,p|\x_0)$
are the Laplace transforms of $G_\infty(\x,t|\x_0)$ and
$j_\infty(\x,t|\x_0)$, respectively.

For the conventional case of a constant reactivity, $\psi(\ell)$ is
given by Eq. (\ref{eq:psi_exp}) and thus $\dtilde{\psi}(\mu) = 1/(1 +
\mu/q)$, so that Eq. (\ref{eq:Gpsi}) is reduced to
\begin{equation}  \label{eq:Gq_Ginf}
\tilde{G}_q(\x,p|\x_0) = \tilde{G}_\infty(\x,p|\x_0)
+ \frac{1}{D} \sum\limits_{k=0}^\infty \frac{[V_k^{(p)}(\x_0)]^* V_k^{(p)}(\x)}{q + \mu_k^{(p)}}  \,.
\end{equation}
Setting $q = 0$, one also gets the identity 
\begin{equation}  \label{eq:G0_Ginf}
\tilde{G}_0(\x,p|\x_0) = \tilde{G}_\infty(\x,p|\x_0)
+ \sum\limits_{k=0}^\infty \frac{[V_k^{(p)}(\x_0)]^* V_k^{(p)}(\x)}{D \mu_k^{(p)}} \,,
\end{equation}
that allows one to write alternative representations
\begin{equation}  \label{eq:Gq_G0}
\tilde{G}_q(\x,p|\x_0) = \tilde{G}_0(\x,p|\x_0)
- \frac{1}{D} \sum\limits_{k=0}^\infty \frac{[V_k^{(p)}(\x_0)]^* V_k^{(p)}(\x)}{\mu_k^{(p)}(1 + \mu_k^{(p)}/q)}  
\end{equation}
and
\begin{align}  \label{eq:Gpsi2}
\tilde{G}_\psi(\x,p|\x_0) & = \tilde{G}_0(\x,p|\x_0) \\  \nonumber
& - \frac{1}{D} \sum\limits_{k=0}^\infty [V_k^{(p)}(\x_0)]^* V_k^{(p)}(\x) \frac{\dtilde{\psi}(\mu_k^{(p)})}{\mu_k^{(p)}}  \,.
\end{align}
As a consequence, we find
\begin{align}  \nonumber
\tilde{j}_\psi(\x,p|\x_0) & = \left. -D\partial_n \tilde{G}_\psi(\x,p|\x_0)\right|_{\pa_R} \\   \label{eq:jpsi}
& = \sum\limits_{k=0}^\infty [V_k^{(p)}(\x_0)]^* v_k^{(p)}(\x) \, \dtilde{\psi}(\mu_k^{(p)}) \,,
\end{align}
where we used the property $(\partial_n V_k^{(p)})|_{\pa_R} =
\mu_k^{(p)} v_k^{(p)}$, given that $v_k^{(p)}$ is an eigenfunction of
the Dirichlet-to-Neumann operator $\M_p$.

In summary, the generalized propagator $G_\psi(\x,t|\x_0)$ and related
quantities fully characterize diffusion-controlled reactions with
irreversible binding described by the probability density
$\psi(\ell)$.  The next step consists in implementing unbinding
events.

\subsection{Unbinding events}
\label{sec:main}

Let us now merge the general mechanism of binding events with a
general waiting time distribution for unbinding events.  Using the
renewal technique, one can compute the propagator $Q(\x,t|\x_0)$ of
reversible diffusion-controlled reactions as
\begin{align} \nonumber
& Q(\x,t|\x_0) = G_\psi(\x,t|\x_0) + \int\limits_{\pa_R} d\x_1 \int\limits_0^t dt_1 \int\limits_{t_1}^t dt'_1 \\  
&  \times  j_\psi(\x_1,t_1|\x_0) \phi(t'_1-t_1) G_\psi(\x,t-t'_1|\x_1) + \ldots .
\end{align}
The first term accounts for the random trajectories from $\x_0$ to
$\x$ without any binding, whose fraction is precisely given by
$G_\psi(\x,t|\x_0)$.  In the second term, the particle binds at time
$t_1$ on the point $\x_1 \in\pa_R$ (with probability
$j_\psi(\x_1,t_1|\x_0) d\x_1 dt_1$), stays in the bound state for the
time $t'_1-t_1$ (with probability $\phi(t'_1-t_1) dt'_1$), unbinds and
diffuses from $\x_1$ to $\x$ within the remaining time $t-t'_1$ (with
probability density $G_\psi(\x,t-t'_1|\x_1)$).  As the binding
position and time as well as the duration of the bound state are
random, one integrates over all their possible realizations.  The next
terms in this expression account for two, three, etc. binding events.

With the help of the Laplace transform, one can turn convolutions in
time into products, yielding
\begin{align}   \label{eq:Qtilde_auxil1}
\tilde{Q}(\x,p|\x_0) & = \tilde{G}_\psi(\x,p|\x_0) \\  \nonumber
& + \int\limits_{\pa_R} d\x_1 \, \tilde{j}_\psi(\x_1,p|\x_0) \tilde{\phi}(p)
\tilde{G}_\psi(\x,p|\x_1) + \ldots .
\end{align}
Using the spectral expansions (\ref{eq:Gpsi}, \ref{eq:jpsi}), one can
evaluate the integrals over the intermediate binding positions $\x_1,
\x_2,\ldots$ in Eq. (\ref{eq:Qtilde_auxil1}) due to the orthogonality of
eigenfunctions $\{v_k^{(p)}\}$ and then sum up all contributions as a
geometric series to get
\begin{align} \nonumber
\tilde{Q}(\x,p|\x_0) & = \tilde{G}_\infty(\x,p|\x_0) 
+ \frac{1}{D} \sum\limits_{k=0}^\infty [V_k^{(p)}(\x_0)]^* V_k^{(p)}(\x) \\   \label{eq:QtildeInf}
& \times \frac{1-\dtilde{\psi}(\mu_k^{(p)})}{\mu_k^{(p)}[ 1 - \tilde{\phi}(p) \dtilde{\psi}(\mu_k^{(p)})]} \,.
\end{align}
Using the identity (\ref{eq:G0_Ginf}), one can rewrite
Eq. (\ref{eq:QtildeInf}) as
\begin{align} \nonumber
\tilde{Q}(\x,p|\x_0) & = \tilde{G}_0(\x,p|\x_0) 
- \frac{1}{D} \sum\limits_{k=0}^\infty [V_k^{(p)}(\x_0)]^* V_k^{(p)}(\x) \\   \label{eq:Qtilde}
& \times \frac{(1-\tilde{\phi}(p))  \, \dtilde{\psi}(\mu_k^{(p)})}{\mu_k^{(p)} [1 - \tilde{\phi}(p) \dtilde{\psi}(\mu_k^{(p)})]}  \,.
\end{align}
This spectral expansion is one of the main general results of the
paper.  It shows how the geometric structure of the confining domain
$\Omega$, expressed in terms of the spectral quantities $\mu_k^{(p)}$
and $V_k^{(p)}(\x)$, is coupled to the binding and unbinding kinetics
expressed in terms of $\dtilde{\psi}(\mu)$ and $\tilde{\phi}(p)$,
respectively.  This expansion holds for any $p\geq 0$ and any
confining domain $\Omega$ with a smooth boundary $\pa$ and a bounded
region $\pa_R$.

In the same way, one can also compute the propagator $Q_b(\x,t|\x_0)$
to be in the bound state at $\x \in \pa_R$ at time $t$.  The renewal
technique yields
\begin{align*}
& Q_b(\x,t|\x_0) = \int\limits_0^t dt' \, j_\psi(\x,t'|\x_0) \, \Phi(t-t') \\
& + \int\limits_{\pa_R} d\x_1 \int\limits_0^t dt_1 \int\limits_{t_1}^t dt'_1 \int\limits_{t'_1}^t dt'\, 
j_\psi(\x,t_1|\x_0) \, \phi(t'_1-t_1) \\
& \times j_\psi(\x,t'-t'_1|\x_1) \Phi(t-t') + \ldots,
\end{align*}
where $\Phi(t) = \int\nolimits_t^{\infty} dt' \, \phi(t')$ is the
probability of no desorption up to time $t$.  The first term describes
the first binding at $\x$ at time $t'$ and staying there until $t$,
while the second and other terms account for multiple
unbinding-rebinding events.  Applying the Laplace transform and using
the spectral expansions, we get
\begin{equation}
\tilde{Q}_b(\x,p|\x_0) = \sum\limits_{k=0}^{\infty} [V_k^{(p)}(\x_0)]^* \, v_k(\x) 
\frac{(1 - \tilde{\phi}(p)) \bar{\psi}(\mu_k^{(p)})}{p[1 - \tilde{\phi}(p)  \bar{\psi}(\mu_k^{(p)})]} \,,
\end{equation}
where we used that $\tilde{\Phi}(p) = (1- \tilde{\phi}(p))/p$.
Evaluating the normal derivative of the propagator
$\tilde{Q}(\x,p|\x_0)$ from Eq. (\ref{eq:Qtilde}) on $\pa_R$, we
deduce
\begin{equation}
p \, \tilde{Q}_b(\x,p|\x_0) = -D \partial_n \tilde{Q}(\x,p|\x_0) \quad (\x\in\pa_R)
\end{equation}
that reads in time domain as
\begin{equation}  \label{eq:BC_Qb}
\partial_t Q_b(\x,t|\x_0) = -D \partial_n Q(\x,t|\x_0)   \quad (\x\in\pa_R).
\end{equation}
In other words, the back-reaction boundary condition
(\ref{eq:Robin_rev2}) remains valid in the general case of
non-Markovian binding/unbinding kinetics.  In turn,
Eq. (\ref{eq:Robin_rev1}) representing the first-order kinetics, is
only retrieved when both $\psi(\ell)$ and $\phi(t)$ are exponential
(see Sec. \ref{sec:bindingM}).

Note that if the particle was initially in the bound state (at a
boundary point $\x_0 \in \pa_R$), one has to include an additional
waiting step until its first desorption.  As a consequence, the
Laplace transform of the associated propagator that we denote
$Q(\x,t|\x_0,b)$ is
\begin{align} \label{eq:Qbtilde}
& \tilde{Q}(\x,p|\x_0,b) = \tilde{\phi}(p) \tilde{Q}(\x,p|\x_0) \\   \nonumber
& = \frac{1}{D} \sum\limits_{k=0}^\infty [V_k^{(p)}(\x_0)]^* V_k^{(p)}(\x)   
 \frac{(1-\dtilde{\psi}(\mu_k^{(p)})) \, \tilde{\phi}(p)}{\mu_k^{(p)}[ 1 - \tilde{\phi}(p) \dtilde{\psi}(\mu_k^{(p)})]} \,,
\end{align}
where we used Eq. (\ref{eq:QtildeInf}).

The integral of Eq. (\ref{eq:Qtilde}) over $\x\in \Omega$ determines
the Laplace transform of the probability $S(t|\x_0)$ that the particle
is unbound at time $t$:
\begin{align}  \label{eq:Stilde}
& \tilde{S}(p|\x_0) = \int\limits_{\Omega} d\x \, \tilde{Q}(\x,p|\x_0) = \frac{1}{p} \\  \nonumber
&  - \sum\limits_{k=0}^\infty [V_k^{(p)}(\x_0)]^* 
\frac{\dtilde{\psi}(\mu_k^{(p)}) (1-\tilde{\phi}(p))}{p [1 - \tilde{\phi}(p) \dtilde{\psi}(\mu_k^{(p)})]} 
\int\limits_{\pa_R} d\x \, v_k^{(p)}(\x) ,
\end{align}
where we used the following identity from \cite{Grebenkov19b}
\begin{equation}
\int\limits_{\Omega} d\x \, V_k^{(p)}(\x) = \frac{D}{p} \mu_k^{(p)} \int\limits_{\pa_R} d\x \, v_k^{(p)}(\x).  
\end{equation}

If there was a uniform concentration $c_0$ of particles at time $0$,
their concentration $c(\x,t)$ at time $t$ is given by
Eq. (\ref{eq:cx_Q}).  According to Eq. (\ref{eq:Qtilde}), the
propagator is symmetric with respect to the exchange of $\x$ and
$\x_0$, $Q(\x,t|\x_0) = Q(\x_0,t|\x)$, so that
\begin{equation}  \label{eq:cx_S}
c(\x,t) = \int\limits_{\Omega} d\x_0 \, c_0\, Q(\x,t|\x_0) = c_0 \, S(t|\x),
\end{equation}
and the latter is given by Eq. (\ref{eq:Stilde}) in the Laplace
domain.  The diffusive flux of particles on the reactive region is
then given in the Laplace domain as
\begin{align}  \label{eq:Jtilde}
\tilde{J}(p) & = \int\limits_{\pa_R} d\x \, \bigl(-D\partial_n \tilde{c}(\x,p)\bigr) \\  \nonumber
& = c_0 D \sum\limits_{k=0}^\infty 
\frac{\mu_k^{(p)} \dtilde{\psi}(\mu_k^{(p)}) (1-\tilde{\phi}(p))}{p [1 - \tilde{\phi}(p) \dtilde{\psi}(\mu_k^{(p)})]} 
\left|\int\limits_{\pa_R} d\x \, v_k^{(p)}(\x) \right|^2 .
\end{align}

In summary, we developed a general mathematical formalism to describe
a very broad class of reversible diffusion-controlled reactions.  At a
microscopic level, one can think of a particle diffusing inside a
confining domain $\Omega$ toward a reactive region $\pa_R$.  When the
particle approached $\pa_R$ close enough, a local short-range
interaction attempts to bind it (e.g., due to attractive electrostatic
potential, formation of covalent or ionic bonds, mutual affinity,
conformational change, local minimum of a potential energy, entropic
barrier, adsorption, transfer through a channel, etc.).  If the
binding attempt fails, the particle continues its bulk diffusion until
the next encounter with $\pa_R$, and so on.  The random number of
failed attempts until the successful binding is represented by the
threshold $\hat{\ell}$ drawn from the probability density $\psi(\ell)$
that characterizes binding kinetics.  After staying in the bound state
for a random waiting time $\hat{t}$ drawn from the probability density
$\phi(t)$ that characterizes unbinding kinetics, the particle is
released into the bulk to resume its diffusion.
As the binding event is preceded by multiple diffusive excursions in
the bulk, the generating function $\dtilde{\psi}(\mu) =
\int\nolimits_0^\infty d\ell \, e^{-\mu\ell} \psi(\ell)$ of the
random threshold $\hat{\ell}$ is tightly coupled to the geometric
structure of the confining domain $\Omega$ that are captured via the
eigenmodes of the Dirichlet-to-Neumann operator.  In turn, the
unbinding event is simply a halt at the binding point for a random
waiting time, which is described by $\phi(t)$ or $\tilde{\phi}(p) =
\int\nolimits_0^\infty dt \, e^{-pt} \phi(t)$, independently of the
confinement.  For more sophisticated unbinding events (such as, e.g.,
surface diffusion), $\tilde{\phi}(p)$ is expected to become coupled to
the geometric structure of the reactive region $\pa_R$.  In this way,
one can potentially retrieve and further generalize the description of
intermittent diffusions \cite{Benichou11}, in particular, those with
alternating tours of bulk and surface diffusion
\cite{Chechkin09,Benichou10b,Benichou11b,Rojo11,Chechkin11,Chechkin12,Rupprecht12a,Rupprecht12b,Berezhkovskii15,Berezhkovskii17}.
However, such an extension is beyond the scope of this paper.

\subsection{Markovian binding kinetics}
\label{sec:bindingM}

Let us first inspect the Markovian binding kinetics determined by the
exponential distribution in Eq. (\ref{eq:psi_exp}).  In this case,
Eq. (\ref{eq:Qtilde}) reads
\begin{equation} \label{eq:Qtilde_Markovian}
\tilde{Q}(\x,p|\x_0) = \tilde{G}_0(\x,p|\x_0) 
- \frac{1}{D} \sum\limits_{k=0}^\infty \frac{[V_k^{(p)}(\x_0)]^* V_k^{(p)}(\x)}{\mu_k^{(p)} (1 + \mu_k^{(p)}/q_p)}  \,,
\end{equation}
with
\begin{equation}  \label{eq:qp_def}
q_p = q (1 - \tilde{\phi}(p)) = \tilde{\kappa}(p)/D \,,
\end{equation}
or, equivalently,
\begin{equation} \label{eq:Qtilde_Markovian2}
\tilde{Q}(\x,p|\x_0) = \tilde{G}_\infty(\x,p|\x_0) 
+ \frac{1}{D} \sum\limits_{k=0}^\infty \frac{[V_k^{(p)}(\x_0)]^* V_k^{(p)}(\x)}{\mu_k^{(p)} + q_p}  \,.
\end{equation}
In line with the former description of reversible reactions in
Sec. \ref{sec:intro}, we retrieved the spectral expansion
(\ref{eq:Gq_G0}), in which the constant reactivity parameter $q$ is
replaced by the $p$-dependent function $q_p$ from
Eq. (\ref{eq:qp_def}).  In other words, the propagator
$\tilde{Q}(\x,p|\x_0)$ satisfies the Robin boundary condition
\begin{equation}  \label{eq:Q_Robin}
-D \partial_n \tilde{Q}(\x,p|\x_0) = \tilde{\kappa}(p) \tilde{Q}(\x,p|\x_0)  \quad (\x \in \pa_R), 
\end{equation}
with $\tilde{\kappa}(p) = q_p D$ (see Sec. \ref{sec:Robin} for further
discussions).  One also sees that the results for irreversible binding
are retrieved by formally setting $\tilde{\phi}(p) \equiv 0$.

The inverse Laplace transform of Eq. (\ref{eq:Q_Robin}) yields a
convolution-type boundary condition on $\pa_R$:
\begin{equation}
- D\partial_n Q(\x,t|\x_0) = \int\limits_0^t dt' \, \K(t-t')\, Q(\x,t'|\x_0),
\end{equation}
with
\begin{equation}
\K(t) = \kappa_0 \bigl(\delta(t) - \phi(t)\bigr),  \quad \kappa_0 = qD.
\end{equation}
This is a generalization of Eqs. (\ref{eq:Robin_conv},
\ref{eq:kappa_exp}).  This is also an extension of the results by
Agmon and Weiss \cite{Agmon89} to arbitrary confining domains.
Whatever the distribution of waiting times is, the memory kernel
induces a sort of delayed feedback from the particles that are
``stored'' in the bound state and progressively released at later
times.  We inspect the consequences of these memory effects for
general binding/unbinding kinetics in Sec. \ref{sec:long}.

\subsection{Robin boundary condition}
\label{sec:Robin}

Let us return to the general setting.  It is easy to check that the
Laplace-transformed propagator $\tilde{Q}(\x,p|\x_0)$ given by
Eq. (\ref{eq:Qtilde}) satisfies the modified Helmholtz equation, as
the conventional propagators do:
\begin{equation}
(p - D \Delta) \tilde{Q}(\x,p|\x_0) = \delta(\x-\x_0)  \quad (\x\in\Omega),
\end{equation}
where $\delta(\x-\x_0)$ is the Dirac distribution.  Expectedly, its
inverse Laplace transform yields the diffusion equation in time
domain,
\begin{equation}
\partial_t Q(\x,t|\x_0) = D \Delta Q(\x,t|\x_0)   \quad (\x\in\Omega),
\end{equation}
with the standard initial condition $Q(\x,0|\x_0) = \delta(\x-\x_0)$
that affirms that $\x_0$ is the starting point at time $0$.  As there
is no surface reaction on the reflecting boundary $\pa_N$, the Neumann
boundary condition applies:
\begin{equation}
\partial_n \tilde{Q}(\x,p|\x_0) = 0 \quad (\x\in\pa_N)
\end{equation}
(and the same holds in the time domain).  Let us now inspect the
boundary condition on the reactive part $\pa_R$.

As we discussed earlier, the Markovian binding kinetics is tightly
related to the Robin boundary condition for the Laplace-transformed
propagator $\tilde{Q}(\x,p|\x_0)$.  However, this boundary condition
does not hold in general for non-Markovian binding.  Using
Eqs. (\ref{eq:QtildeInf}, \ref{eq:Qtilde}), one can easily check that
\begin{align*}
\left. -D \partial_n \tilde{Q}(\x,p|\x_0)\right|_{\x\in \pa_R} & = \sum\limits_{k=0}^\infty [V_k^{(p)}(\x_0)]^* V_k^{(p)}(\x) \\
& \times \frac{\dtilde{\psi}(\mu_k^{(p)}) (1 - \tilde{\phi}(p))}{1 - \tilde{\phi}(p) \dtilde{\psi}(\mu_k^{(p)})} \,, \\
\left. D \tilde{Q}(\x,p|\x_0)\right|_{\x\in \pa_R} & = \sum\limits_{k=0}^\infty [V_k^{(p)}(\x_0)]^* V_k^{(p)}(\x) \\
& \times \frac{1 - \dtilde{\psi}(\mu_k^{(p)})}{\mu_k^{(p)}[1 - \tilde{\phi}(p) \dtilde{\psi}(\mu_k^{(p)})]} \,.
\end{align*}
These two functions can satisfy the Robin boundary condition
(\ref{eq:Q_Robin}) only if each term in one series is proportional to
the corresponding term in the other, i.e., if
\begin{equation*}
\frac{\dtilde{\psi}(\mu_k^{(p)}) (1 - \tilde{\phi}(p))}{1 - \tilde{\phi}(p) \dtilde{\psi}(\mu_k^{(p)})} 
= q_p \frac{1 - \dtilde{\psi}(\mu_k^{(p)})}{\mu_k^{(p)}[1 - \tilde{\phi}(p) \dtilde{\psi}(\mu_k^{(p)})]} 
\end{equation*}
for some proportionality coefficient $q_p$, from which
\begin{equation}  \label{eq:qp_general}
q_p = (1 - \tilde{\phi}(p)) \frac{\mu_k^{(p)} \dtilde{\psi}(\mu_k^{(p)})}{1 - \dtilde{\psi}(\mu_k^{(p)})} \,.
\end{equation}
As the left-hand side does not depend on $k$, the right-hand side
should not as well, i.e., one should have
\begin{equation*}
\frac{\mu_k^{(p)} \dtilde{\psi}(\mu_k^{(p)})}{1 - \dtilde{\psi}(\mu_k^{(p)})} = q 
\end{equation*}
for some constant $q$, and thus $\dtilde{\psi}(\mu_k^{(p)}) = 1/(1 +
\mu_k^{(p)}/q)$, i.e., the random threshold should obey an exponential
distribution with the rate $q$.  In other words, if the distribution
of $\hat{\ell}$ is exponential, the propagator $\tilde{Q}(\x,p|\x_0)$
satisfies the Robing boundary condition (\ref{eq:Q_Robin}).  In turn,
any non-exponential density $\psi(\ell)$ would in general invalidate
the Robin boundary condition.

Despite this general statement, the Robin boundary condition can still
re-appear in some situations.  For instance, let us look at the
concentration $c(\x,t)$ with the uniform initial condition $c_0(\x_0)
= c_0$, which is proportional to $S(t|\x)$ according to
Eq. (\ref{eq:cx_S}).  In some symmetric domains, the integral over
$\x\in\pa_R$ may cancel all contributions in Eq. (\ref{eq:Stilde})
except for $k = 0$.  In this specific case, $\tilde{c}(\x,p)$ would
satisfy the Robin boundary condition
\begin{equation}  \label{eq:S_Robin}
- \partial_n \tilde{c}(\x,p) = q_p \, \tilde{c}(\x,p)  \quad (\x\in \pa_R),
\end{equation}
with $q_p$ given by Eq. (\ref{eq:qp_general}) for $k = 0$.  Indeed, if
both $\tilde{c}(\x,p)$ and $\partial_n \tilde{c}(\x,p)$ (restricted on
$\pa_R$) are determined by a single eigenmode, they are necessarily
proportional to each other.  In Sec. \ref{sec:sphere}, we discuss an
explicit example of a spherical target, whose rotational symmetry
leads to this situation.

The inverse Laplace transform of Eq. (\ref{eq:S_Robin}) yields again a
convolution-type Robin boundary condition (\ref{eq:Robin_conv}) in
time domain, with the memory kernel $\K(t)$ obtained by the inverse
Laplace transform of $\tilde{\kappa}(p) = q_p D$:
\begin{equation}
\K(t) = \L^{-1} \left\{ D (1 - \tilde{\phi}(p)) \frac{\mu_0^{(p)} \dtilde{\psi}(\mu_0^{(p)})}{1 - \dtilde{\psi}(\mu_0^{(p)})} \right\}.
\end{equation}
In this case, the statistics of both binding and unbinding events,
incorporated through $\dtilde{\psi}(\mu_0^{(p)})$ and
$\tilde{\phi}(p)$, as well as the geometric structure of the domain,
represented by $\mu_0^{(p)}$, all determine the memory kernel.  This
point was first mentioned in
\cite{Grebenkov20}.  Moreover, Bressloff showed for simple 
one-dimensional settings that such memory kernels can be heavy-tailed
\cite{Bressloff22a,Bressloff23a,Bressloff23b}.  We return to this
point in Sec. \ref{sec:sphere} for the case of a spherical reactive
region.

\subsection{Long-time behavior}
\label{sec:long}

To conclude this section, we investigate the long-time behavior of the
propagator $Q(\x,t|\x_0)$.  We consider general asymptotic expressions
\begin{align}
\dtilde{\psi}(\mu) & \approx 1 - \mu^\alpha \ell_0^\alpha + \ldots  \quad (\mu \to 0), \\  \label{eq:phi_p0}
\tilde{\phi}(p) & \approx 1 - p^\beta \ddelta^\beta + \ldots \quad (p\to 0), 
\end{align}
with some exponents $0 < \alpha \leq 1$ and $0 < \beta \leq 1$ and
length and time scales $\ell_0$ and $\ddelta$ that characterize
binding and unbinding events, respectively.  When $\beta = 1$, the
mean waiting time is finite and equal to $\ddelta$.  Similarly, when
$\alpha = 1$, the mean threshold is finite and equal to $\ell_0$.
Using the asymptotic expansion \cite{Grebenkov19a}
\begin{equation}  \label{eq:mu0_p0}
\mu_0^{(p)} \approx \frac{|\Omega|}{D|\pa_R|} p + \ldots  \quad (p\to 0),
\end{equation}
where $|\Omega|$ is the volume of $\Omega$, we can write
\begin{equation}  \label{eq:psitilde_0}
\dtilde{\psi}(\mu_0^{(p)}) \approx 1 - p^\alpha \ttau^\alpha + \ldots  \quad (p\to 0),
\end{equation}
where
\begin{equation}  \label{eq:tau}
\ttau = \frac{\ell_0 |\Omega|}{D|\pa_R|} 
\end{equation}
is a time scale associated to binding events (see Appendix
\ref{sec:binding_time}).
Finally, we note that $v_0^{(0)} = 1/\sqrt{|\pa_R|}$ and
thus
\begin{equation}
V_0^{(0)}(\x_0) = \int\limits_{\pa_R} d\x \, \tilde{j}_\infty(\x,0|\x_0) v_0^{(0)}(\x) = \frac{1}{\sqrt{|\pa_R|}}
\end{equation}
due to the normalization of the probability flux density
$j_\infty(\x,t|\x_0)$.  

The first term in Eq. (\ref{eq:Qtilde}) behaves as
$\tilde{G}_0(\x,p|\x_0) \approx 1/(p|\Omega|)$ in the limit $p\to 0$,
ensuring that the conventional propagator $G_0(\x,t|\x_0)$ approaches
the uniform distribution in the bounded domain $\Omega$ with
reflecting boundary $\pa$:
\begin{equation}
G_0(\x,t|\x_0) \approx \frac{1}{|\Omega|} + O(exp)  \quad (t\to \infty),
\end{equation}
where $O(exp)$ denotes an exponentially decaying correction.  In the
second term of Eq. (\ref{eq:Qtilde}), let us denote the contributions
as
\begin{equation}
\tilde{Q}_k(p) = \frac{1}{D} [V_k^{(p)}(\x_0)]^* V_k^{(p)}(\x) \frac{\dtilde{\psi}(\mu_k^{(p)}) (1-\tilde{\phi}(p))}
{\mu_k^{(p)} [1 - \tilde{\phi}(p) \dtilde{\psi}(\mu_k^{(p)})]} \,.
\end{equation}
For the ground eigenmode $k = 0$, the substitution of the above
small-$p$ expansions yields
\begin{equation}  \label{eq:Q0p}
\tilde{Q}_0(p) \approx \frac{p^{\beta-1} \ddelta^\beta}
{|\Omega|\bigl(p^\alpha \ttau^\alpha + p^\beta \ddelta^\beta \bigr)}   \quad (p\to 0).
\end{equation}
Expectedly, it does not depend on the starting point $\x_0$, nor on
the arrival point $\x$.  In turn, for other eigenmodes with $k > 0$,
one has $\mu_k^{(p)} \to \mu_k^{(0)} > 0$ as $p\to 0$ and therefore
\begin{equation}
\tilde{Q}_k(p) \approx  \frac{[V_k^{(0)}(\x_0)]^* V_k^{(0)}(\x) \ddelta^\beta  \dtilde{\psi}(\mu_k^{(0)})}
{D \mu_k^{(0)} \bigl(1 - \dtilde{\psi}(\mu_k^{(0)})\bigr)} p^{\beta}  \quad (p\to 0).
\end{equation}
For any combination of the exponents $\alpha$ and $\beta$, the inverse
Laplace transform of this term yields a subleading contribution as
compared to $Q_0(t)$.  As a consequence, we focus on the leading-order
term $Q_0(t)$.

We distinguish four regimes.

(i) If $\alpha = \beta = 1$, one has in the leading order
\begin{equation}
\tilde{Q}_0(p) \approx \frac{\ddelta}{|\Omega|(\ddelta + \ttau)p }  \quad (p\to 0),
\end{equation}
so that
\begin{equation}
Q(\x,t|\x_0) \approx \frac{1}{|\Omega| (1 + \ddelta/\ttau)} + O(exp)  \quad (t\to \infty).
\end{equation}
Expectedly, when both binding and unbinding events are characterized
by finite means, the system evolves toward an equilibrium
distribution, which is uniform inside the confining domain.  The
probability to be in the unbound state is $1/(1 + \ddelta/\ttau)$.  An
exponential relaxation to the equilibrium is the reminiscent feature
of restricted diffusion in a bounded confining domain.  It is
drastically different from diffusion in unbounded domains, for which
the particle can diffuse arbitrarily far away from the reactive region
so that the duration of each bulk exploration between consecutive
bindings can be very long; in particular, the survival probability for
irreversible reactions and the probability to be in the bound state
for reversible reactions are known to exhibit power-law decays in time
\cite{Redner,Agmon84,Agmon89,Levitz08,Bray13,Levernier19,Grebenkov21}.

(ii) If $\alpha < \beta \leq 1$, one has in the leading order
\begin{equation}
\tilde{Q}_0(p) \approx \frac{\ddelta^\beta}{|\Omega|\ttau^\alpha} p^{-(1+\alpha-\beta)}    \quad (p\to 0),
\end{equation}
so that
\begin{equation}
Q(\x,t|\x_0) \approx \frac{1}{|\Omega|}
\biggl(1 - \frac{\ddelta^\beta \, t^{\alpha-\beta}}{\ttau^\alpha \Gamma(1+\alpha-\beta)} \biggr)   \quad (t\to \infty).
\end{equation}
As the binding events are characterized by the smaller exponent
$\alpha$, unbinding events more probable at long times.  As a
consequence, the particle is asymptotically unbound but the approach
to this regime is controlled by a slow power-law decay, with the
exponent $\alpha-\beta$ (much faster decaying exponential corrections
are not relevant here and thus omitted).

(iii) If $\beta < \alpha \leq 1$, one has in the leading order
\begin{equation}
\tilde{Q}_0(p) \approx \frac{1}{|\Omega| p} \biggl(1 - \frac{\ttau^\alpha}{\ddelta^\beta} p^{\alpha-\beta}\biggr)
 \quad (p\to 0),
\end{equation}
so that
\begin{equation}
Q(\x,t|\x_0) \approx \frac{\ttau^\alpha  \, t^{\beta-\alpha}}{|\Omega|\ddelta^\beta \Gamma(1+\beta-\alpha)} 
   \quad (t\to \infty).
\end{equation}
In this case, binding events are characterized by the larger exponent
$\alpha$ and thus more probable than the unbinding events so that the
particle tends to be asymptotically in the bound state.  In
particular, the probability to be in the unbound state slowly vanishes
as $t^{\beta-\alpha}$, with the exponent $\beta-\alpha$.

(iv) If $\alpha = \beta < 1$, we have
\begin{equation}
\tilde{Q}_0(p) \approx \frac{1}{|\Omega| p(1 + (\ttau/\ddelta)^\alpha)} \biggl(1 
- \frac{\ttau^{2\alpha} \, p^{\alpha}}{\ttau^\alpha + \ddelta^\alpha} \biggr)   \quad (p\to 0),
\end{equation}
so that
\begin{align}   \label{eq:Qt_asympt4}
Q(\x,t|\x_0) & \approx \frac{1}{|\Omega|(1 + (\ddelta/\ttau)^\alpha)} \\  \nonumber
& \times \biggl(1 + \frac{\ddelta^\alpha \, t^{-\alpha}}{(1+(\ddelta/\ttau)^\alpha) \Gamma(1-\alpha)} \biggr) 
\quad (t\to \infty) .
\end{align}
In this subtle setting, the occurrence of binding and unbinding events
is equally rare at long times so that the system approaches an
equilibrium uniform distribution, but this power-law approach is much
slower than in the case $\alpha = \beta = 1$.  The asymptotic
probability of the unbound state is $1/(1 + (\ddelta/\ttau)^\alpha)$.

The integral over $\x\in\Omega$ of the above relations removes the
factor $1/|\Omega|$ and yields the long-time asymptotic behavior of
the probability $S(t|\x_0)$, while its multiplication by $c_0$
determines the asymptotic behavior of the concentration $c(\x,t)$.

\section{Spherical target}
\label{sec:sphere}

To illustrate the above general results, we consider restricted
diffusion between two concentric spheres of radii $R$ and $L$: $\Omega
= \{\x\in\R^3 ~:~ R < |\x| < L\}$.  The inner sphere represents the
reactive region $\pa_R$, while the outer sphere is the reflecting
boundary $\pa_N$.  This emblematic model for diffusion-controlled
reactions was thoroughly investigated (see \cite{Grebenkov18,Reva21}
and references therein).  In a similar way, one can obtain the exact
results for restricted diffusion on an interval, a half-line, a
circular annulus, and the exterior of the cylinder (see
\cite{Grebenkov20b} and Appendix \ref{sec:interval}).  For the sake of
clarity, we focus on the probability $S(t|\x_0)$ (or, equivalently, on
the concentration $c(\x,t)$) and on the diffusive flux $J(t)$.

The eigenbasis of the Dirichlet-to-Neumann operator in this domain is
known \cite{Grebenkov20b}.  In fact, the eigenfunctions $v_k^{(p)}$
are given by the normalized spherical harmonics so that the integral
over $\pa_R$ in Eq. (\ref{eq:Stilde}) cancels all terms except the
ground eigenmode, for which $v_0^{(p)}(\x) = 1/\sqrt{|\pa_R|}$.  We
get then
\begin{equation}  \label{eq:Stilde_sphere}
\tilde{S}(p|\x_0) = \frac{1}{p} - g_0^{(p)}(r_0) 
\frac{\dtilde{\psi}(\mu_0^{(p)}) (1-\tilde{\phi}(p))}{p [1 - \tilde{\phi}(p) \dtilde{\psi}(\mu_0^{(p)})]}  \,,
\end{equation}
where $r_0 = |\x_0|$,
\begin{subequations}
\begin{align}   \label{eq:g0}
g_0^{(p)}(r) &= \frac{R}{r} \, \frac{w(L\sqrt{p/D}, r\sqrt{p/D})}{w(L\sqrt{p/D}, R\sqrt{p/D})} \,, \\
w(x,y) & = \sinh(x-y) - x \cosh(x-y) ,
\end{align}
\end{subequations}
and
\begin{equation}  \label{eq:mu0_sphere}
\mu_0^{(p)} = \frac{1}{R} + \sqrt{p/D} \, \frac{L\sqrt{p/D} \tanh((L-R)\sqrt{p/D}) - 1}{L\sqrt{p/D} - \tanh((L-R)\sqrt{p/D})} \,.
\end{equation}
The diffusive flux (\ref{eq:Jtilde}) reads in the Laplace domain:
\begin{equation}   \label{eq:Jtilde_sphere}
\tilde{J}(p) = 4\pi R^2 c_0 D \frac{\mu_0^{(p)} (1-\tilde{\phi}(p)) \dtilde{\psi}(\mu_0^{(p)})}
{p[1 - \tilde{\phi}(p) \dtilde{\psi}(\mu_0^{(p)})]} \,.
\end{equation} 
If the particle was initially in the bound state, the inclusion of an
additional waiting step yields
\begin{equation}
\tilde{S}(p|b) = \tilde{\phi}(p) \, \tilde{S}(p|\x_0)|_{\pa_R} 
 = \frac{\tilde{\phi}(p) (1 - \dtilde{\psi}(\mu_0^{(p)}))}{p [1 - \tilde{\phi}(p) \dtilde{\psi}(\mu_0^{(p)})]} \,,
\end{equation}
which is independent of the starting point $\x_0$ due to the
rotational invariance of the domain. 

For Markovian binding kinetics with $\psi(\ell)$ from
Eq. (\ref{eq:psi_exp}), one has $\dtilde{\psi}(\mu) = 1/(1 + \mu/q)$,
so that the above expressions are reduced to
\begin{equation}
\tilde{S}(p|\x_0) = \frac{1}{p} -  \frac{g_0^{(p)}(r_0)[1-\tilde{\phi}(p)]}{p[1 - \tilde{\phi}(p) + \mu_0^{(p)}/q]} \,,
\end{equation}
\begin{equation}
\tilde{S}(p|b) = \frac{\tilde{\phi}(p)}{p [1 +q (1- \tilde{\phi}(p))/\mu_0^{(p)}]} \,,
\end{equation}
and
\begin{equation}  \label{eq:Jp_sphere}
\tilde{J}(p) = 4\pi R^2 c_0 D \frac{\mu_0^{(p)} (1-\tilde{\phi}(p))}
{p(1 - \tilde{\phi}(p) + \mu_0^{(p)}/q)} \,.
\end{equation}

To show the behavior of these quantities in time domain, we will
compute their inverse Laplace transforms numerically by using the
Talbot algorithm \cite{Talbot79}.

\subsection{Mittag-Leffler model}

\begin{figure}[t!]
\begin{center}
\includegraphics[width=85mm]{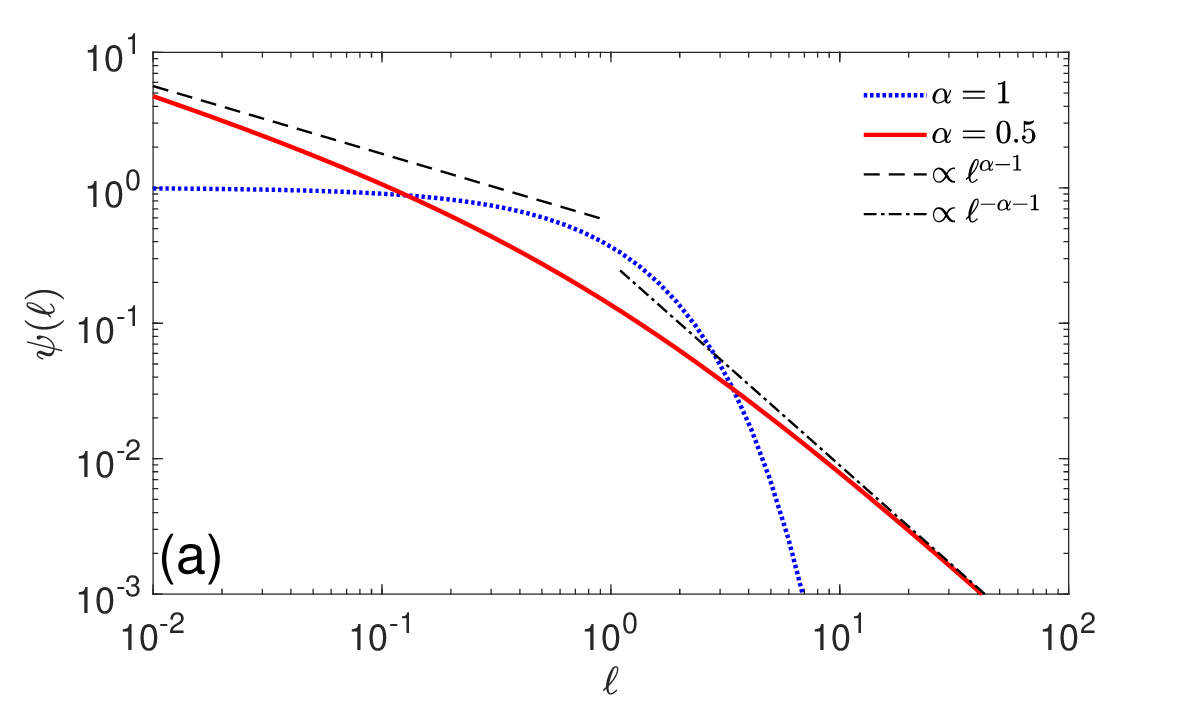} 
\includegraphics[width=85mm]{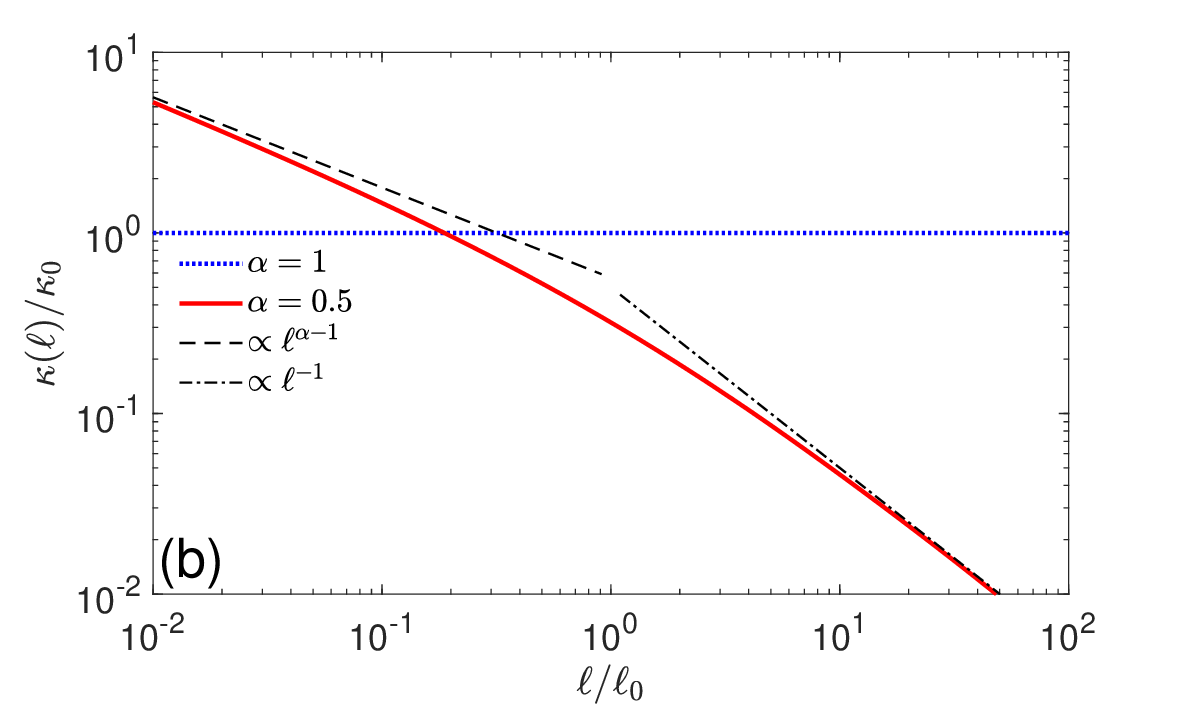} 
\end{center}
\caption{
{\bf (a)} The Mittag-Leffler probability density $\psi(\ell)$ from
Eq. (\ref{eq:psi_ML}), with $\ell_0 = 1$ and two exponents: $\alpha =
1$ and $\alpha = 0.5$.  Thin lines indicate the asymptotic relations
(\ref{eq:ML_asympt}).  {\bf (b)} The associated encounter-dependent
reactivity $\kappa(\ell)$ given by Eq. (\ref{eq:kappa_ML}).  Thin
lines indicate the asymptotic relations (\ref{eq:kappa_ML_asympt}).}
\label{fig:psi}
\end{figure}

In order to illustrate various features of non-Markovian
binding/unbinding kinetics, we introduce the Mittag-Leffler model, in
which both binding and unbinding events are characterized by
Mittag-Leffler distributions:
\begin{equation}  \label{eq:psi_ML}
\psi(\ell) = -E_{\alpha,0}(-(\ell/\ell_0)^\alpha)/\ell 
\end{equation}
for the threshold $\hat{\ell}$ determining each binding event, and
\begin{equation}  \label{eq:phi_ML}
\phi(t) = - E_{\beta,0}(-(t/\ddelta)^\beta)/t 
\end{equation}
for the waiting time $\hat{t}$ in each bound state, where
\begin{equation}
E_{\gamma,\delta}(z) = \sum\limits_{k=0}^\infty \frac{z^k}{\Gamma(\gamma k + \delta)} 
\end{equation}
is the Mittag-Leffler function.  The associated generating 
functions are particularly simple: $\dtilde{\psi}(\mu) = 1/(1 + (\mu
\ell_0)^\alpha)$ and $\tilde{\phi}(p) = 1/(1 + (p\ddelta)^\beta)$.
Since $E_{1,0}(z) = z e^{z}$, the above distributions become
exponential (with $q = 1/\ell_0$ and $\koff = 1/t_d$) when $\alpha =
1$ or $\beta = 1$.  When the exponent is smaller than $1$, the
probability density exhibits a power-law behavior in both limits,
e.g.,
\begin{subequations}  \label{eq:ML_asympt}
\begin{align}   \label{eq:ML_asympt0}
\psi(\ell) & \approx \frac{\ell^{\alpha-1}}{\ell_0^\alpha \Gamma(\alpha)}  \quad (\ell\to 0) ,\\  \label{eq:ML_asymptinf}
\psi(\ell) & \approx \frac{\ell_0^\alpha \, \ell^{-\alpha-1}}{|\Gamma(-\alpha)|}  \quad (\ell\to \infty) .
\end{align}
\end{subequations}
Note that if $\alpha$ or $\beta$ is equal to $1/2$, the above
Mittag-Leffler function can be expressed in terms of the error
function $\erf(z)$, e.g., one has for $\alpha = 1/2$
\begin{equation}  \label{eq:phi_ML_12}
\psi(\ell) = \frac{1}{\sqrt{\pi \ell_0 \ell}} - \frac{1}{\ell_0}\, \erfcx(\sqrt{\ell/\ell_0}) ,
\end{equation}
where $\erfcx(z) = e^{z^2}(1-\erf(z))$ is the scaled complementary
error function (see Fig. \ref{fig:psi}(a)).

\subsection{Diffusion toward a sphere in $\R^3$}
\label{sec:sphereInf}

Before dwelling on restricted diffusion in the bounded spherical
domain, it is instructive to have a look at the limit $L\to\infty$
when the outer reflecting boundary is moved away.  This limit
corresponds to the most well-studied setting of a particle diffusing
in $\R^3$ toward a reactive sphere of radius $R$.  This is the problem
that was addressed in the seminal works by Smoluchowski, Collins and
Kimball and many others.  Even though we focused on bounded domains in
Sec. \ref{sec:theory}, the derived spectral expansions generally
remain valid for unbounded domains as well, provided that the reactive
region $\pa_R$ is bounded.  While the mathematical treatment of
general unbounded domains goes beyond the scope of the paper, taking
the limit $L\to \infty$ in the considered example of concentric
spheres is straightforward.  For instance, one gets immediately
\begin{align} \nonumber
\tilde{S}(p|\x_0) & = \frac{1}{p} - \frac{R}{r_0} e^{-(r_0-R)\sqrt{p/D}}  \\    \label{eq:Stilde_sphere_Linf}
& \times \frac{(1-\tilde{\phi}(p)) \dtilde{\psi}(1/R + \sqrt{p/D})}{p [1 - \tilde{\phi}(p) \dtilde{\psi}(1/R + \sqrt{p/D})]}  
\end{align}
and
\begin{equation} 
\tilde{S}(p|b) = \frac{\tilde{\phi}(p) (1-\dtilde{\psi}(1/R + \sqrt{p/D}))}
{p [1 - \tilde{\phi}(p) \dtilde{\psi}(1/R + \sqrt{p/D})]}  \,.
\end{equation}
For the Markovian binding, these expressions are reduced to
\begin{equation}    \label{eq:Sp_sphere_Linf}
1 - p \tilde{S}(p|\x_0) = \frac{R}{r_0}  \, \frac{e^{-(r_0-R)\sqrt{p/D}} (1-\tilde{\phi}(p))}{1- \tilde{\phi}(p) + 
\frac{1/R + \sqrt{p/D}}{q}}  
\end{equation}
and
\begin{equation}  \label{eq:Sb_sphere_Linf}
p \tilde{S}(p|b) = \frac{\tilde{\phi}(p)}{1 + (1- \tilde{\phi}(p)) \frac{\kappa_0 R}{D(1 + R\sqrt{p/D})}}  \,.
\end{equation}
An equation similar to Eq. (\ref{eq:Sp_sphere_Linf}) was earlier
reported in \cite{Kim99,Scher23} for the conventional case of
Markovian unbinding, with $\tilde{\phi}(p) = 1/(1 + p/\koff)$.  In
this particular case, the inverse Laplace transform can be explicitly
inverted to express $S(t|\x_0)$ in terms of error functions
\cite{Kim99}.  Equation (\ref{eq:Sp_sphere_Linf}) turns out to
be a generalization to the non-Markovian setting.  In turn,
Eq. (\ref{eq:Sb_sphere_Linf}) is identical to the expression (2.12)
derived by Agmon and Weiss for this particular setting \cite{Agmon89},
with their notation $k_a = 4\pi R^2 \kappa_0$.

For Markovian binding/unbinding kinetics, Eq. (\ref{eq:Jp_sphere})
yields
\begin{equation}  \label{eq:Jp_sphere_inf}
\tilde{J}(p) = 4\pi R c_0 D \frac{1 + R\sqrt{p/D}}{p[1 + \frac{1}{qR} (1 + R\sqrt{p/D})(1 + \koff/p)]} \,.
\end{equation}
This Laplace transform can be inverted explicitly by finding the roots
of a cubic polynomial in powers of $\sqrt{p/D}$ in the denominator,
expanding it into partial fractions and inverting them (see details in
\cite{Kim99}).  If there is no unbinding kinetics (i.e., $\koff = 0$),
the inverse Laplace transform of this expression yields the diffusive
flux found by Collins and Kimball \cite{Collins49}:
\begin{equation}  \label{eq:Jt_CK}
J(t) = \frac{4\pi R^2 c_0 qD}{1 + qR} \bigg(1 + qR\, \erfcx\bigl(\sqrt{Dt} (1/R + q)\bigr)\biggr).
\end{equation}
In the limit $q\to \infty$ (perfect reactions), one retrieves the
Smoluchowski formula:
\begin{equation}  \label{eq:Jt_Smol}
J(t) = J_{\rm S} \biggl(1 + \frac{R}{\sqrt{\pi Dt}}\biggr),
\end{equation}
where 
\begin{equation}  \label{eq:Jinf}
J_{\rm S} = 4\pi R c_0 D 
\end{equation}
is the Smoluchowski steady-state flux in the long-time limit.  

The crucial difference from the bounded case is the possibility of an
escape to infinity in three dimensions.  From the mathematical point
of view, this possibility is reflected in the strictly positive limit
of the smallest eigenvalue: $\mu_0^{(p)} \to 1/R$ as $p\to 0$, whereas
for bounded domains, one had $\mu_0^{(0)} = 0$.  As discussed in
\cite{Grebenkov19a}, the mean boundary local time approaches a
finite limit so that the particle undertakes a finite number of
binding events before escaping to infinity.  If the mean waiting time
in each bound state is finite, the mean value of the overall duration
of binding events is also finite, and this mean determines the
characteristic time scale $T$, above which the probability $S(t|\x_0)$
approaches $1$ exponentially fast, i.e., $1 - S(t|\x_0) \propto
e^{-t/T}$.  In turn, if the mean waiting time is infinite, one can
insert Eq. (\ref{eq:phi_p0}) with $\beta < 1$ into
Eq. (\ref{eq:Stilde_sphere_Linf}) to get in the leading order
\begin{equation}  
\tilde{S}(p|\x_0) \approx \frac{1}{p} - \frac{R \ddelta^\beta \, p^{\beta-1}}{r_0(1/\dtilde{\psi}(1/R)-1)}  
\quad (p\to 0) ,
\end{equation}
that yields 
\begin{equation}
S(t|\x_0) \approx 1 - \frac{R \, (t/\ddelta)^{-\beta}}{r_0(1/\dtilde{\psi}(1/R)-1)\Gamma(1-\beta)} 
\quad (t\to \infty).
\end{equation}
Expectedly, the slow, power-law approach of the probability
$S(t|\x_0)$ to $1$ is controlled by long halts in the bound states.
The role of $\dtilde{\psi}(1/R)$ that determines the amplitude in
front of the power law, was revealed in \cite{Grebenkov20}.  For
instance, in the case of conventional binding in
Eq. (\ref{eq:psi_exp}), one has $\dtilde{\psi}(1/R) = 1/(1 + 1/(qR))$
so that
\begin{equation}
S(t|\x_0) \approx 1 - \frac{qR^2 \, (t/\ddelta)^{-\beta}}{r_0\Gamma(1-\beta)}   \quad (t\to \infty).
\end{equation}
One sees that a highly reactive region (large $q$) ensures rapid
binding events and therefore a larger amplitude of the correction
term, i.e., a slower approach to $1$.

\subsection{Probability to be in the unbound state}

From now on, we return to restricted diffusion between concentric
spheres.  In the following, we fix the units of length and time by
setting $R = 1$ and $D = 1$.

\begin{figure}[t!]
\begin{center}
\includegraphics[width=85mm]{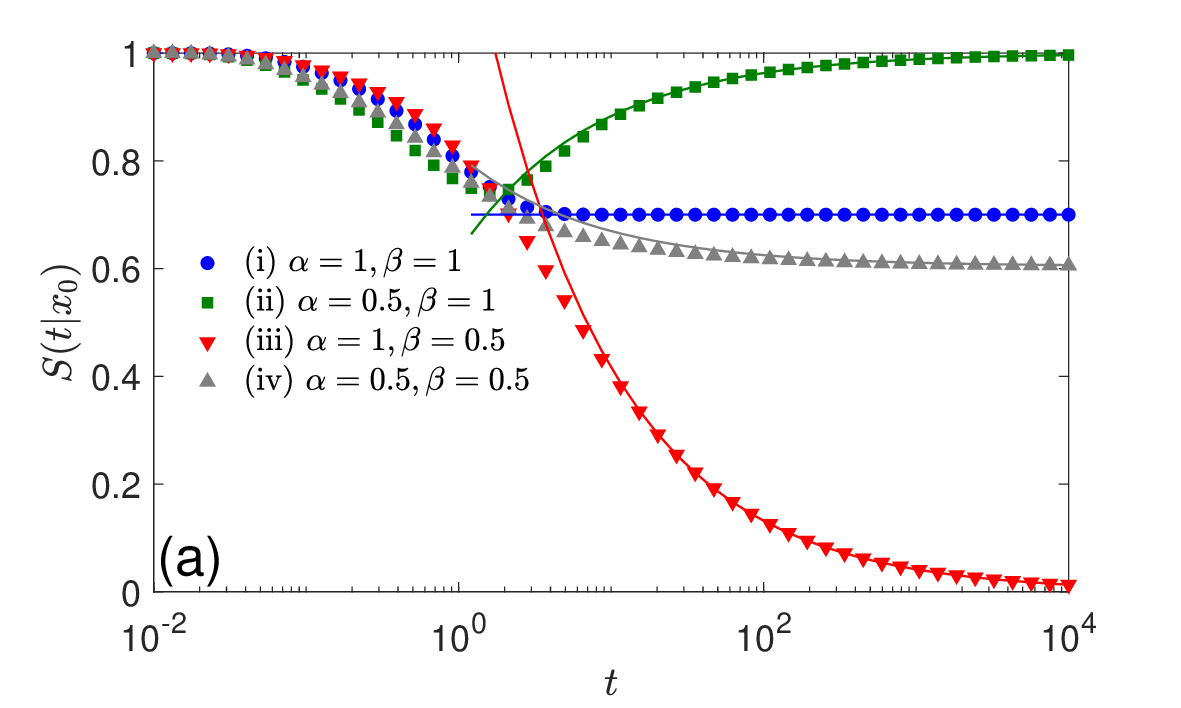} 
\includegraphics[width=85mm]{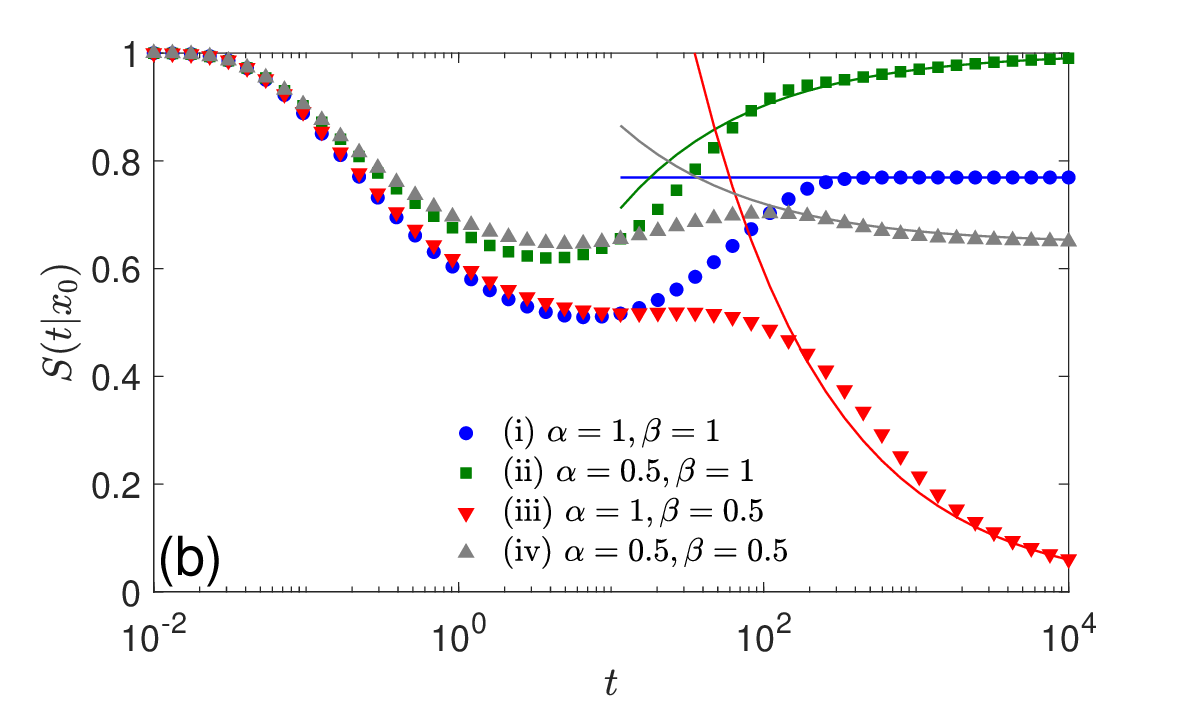} 
\end{center}
\caption{
The probability $S(t|\x_0)$ of finding the particle in the unbound
state at time $t$, for diffusion between concentric spheres, with $D =
1$, $R = 1$, $|\x_0| = 1.5$, $L = 2$ {\bf (a)} and $L = 10$ {\bf (b)}.
The random threshold and waiting time obey Mittag-Leffler
distributions (\ref{eq:psi_ML}, \ref{eq:phi_ML}), with $\ell_0 = 1$
and $\ddelta = 1$ {\bf (a)} and $\ell_0 = 0.1$ and $\ddelta = 10$ {\bf
(b)}, and four combinations of the exponents $\alpha$ and $\beta$ as
indicated in the legend.  Symbols present the inverse Laplace
transform of the exact expression (\ref{eq:Stilde_sphere}) obtained by
the Talbot algorithm, whereas lines indicate the long-time asymptotic
relations discussed in Sec. \ref{sec:long}.  The values of $\ell_0$
and $\ddelta$ were changed for the panel {\bf (b)} for a better
visualization.}
\label{fig:Qt}
\end{figure}

Figure \ref{fig:Qt} illustrates the behavior of the probability
$S(t|\x_0)$ of finding a particle in the unbound state in four
different regimes discussed in Sec. \ref{sec:long}.  Two panels
correspond to strong and weak confinements, with $L = 2$ and $L = 10$,
respectively.  In both cases, the asymptotic relations derived in
Sec. \ref{sec:long} accurately capture the long-time behavior of
$S(t|\x_0)$.  For the regimes (i), (iii) and (iv), $S(t|\x_0)$
exhibits a monotonous decrease in panel (a); in contrast, this
function achieves a minimum in the regime (ii).  This minimum is
necessarily present for any setting because $S(t|\x_0)$ starts from
$1$ at $t=0$ and returns to $1$ in the limit $t\to\infty$.  In turn,
the monotonous decrease of $S(t|\x_0)$ in other regimes is specific to
the chosen set of parameters.  For instance, the panel (b) illustrates
the case $L = 10$, for which $S(t|\x_0)$ is not monotonous in all
regimes.  Moreover, a local maximum is observed for the case (iv).
Changing the parameters $\ell_0$ and $\ddelta$ (as well as $L$ and
$|\x_0|$), one can achieve other situations (not shown), in which this
local maximum is either enhanced or removed.  To rationalize its
origin, let us revise the short-time and long-time behaviors of the
probability $S(t|\x_0)$.  At short times, the binding kinetics is the
limiting factor so that $S(t|\x_0)$ is almost independent of the
unbinding kinetics (see Appendix \ref{sec:Sshort_sphere} for technical
details).  In particular, a rapid decrease of $S(t|\x_0)$ can be
achieved either by increasing the ``reactivity'' of $\pa_R$ (i.e.,
taking smaller $\ell_0$ or $t_a$) or by choosing the starting position
$\x_0$ closer to $\pa_R$.  After this initial drop, unbinding events
come in play and start to ``compete'' with binding events to slowly
reach an equilibrium limit $S(\infty|\x_0) = 1/(1 +
(t_d/t_a)^\alpha)$.  Changing the timescale $t_d$ of unbinding events
allows one to control the value of this limit, independently of the
short-time behavior.  Moreover, as the second term in
Eq. (\ref{eq:Qt_asympt4}) is positive, the limit $S(\infty|\x_0)$ is
approached from above.  As a consequence, if the initial drop led to
intermediate values of $S(t|\x_0)$ that are below $S(\infty|\x_0)$, a
local maximum should be present.  In a similar way, one can
rationalize the presence of a local maximum for the case (iii) that we
observed for a different set of parameters (not shown).

\subsection{Diffusive flux}

In order to quantify the uptake on the reactive region, we consider
the diffusive flux $J(t)$, which is obtained via a numerical inversion
of the Laplace transform in Eq. (\ref{eq:Jtilde_sphere}).  To grasp
the main features of this quantity, we start with the conventional
case of Markovian binding and unbinding kinetics (i.e., $\alpha =
\beta =1$).

\begin{figure}[t!]
\begin{center}
\includegraphics[width=85mm]{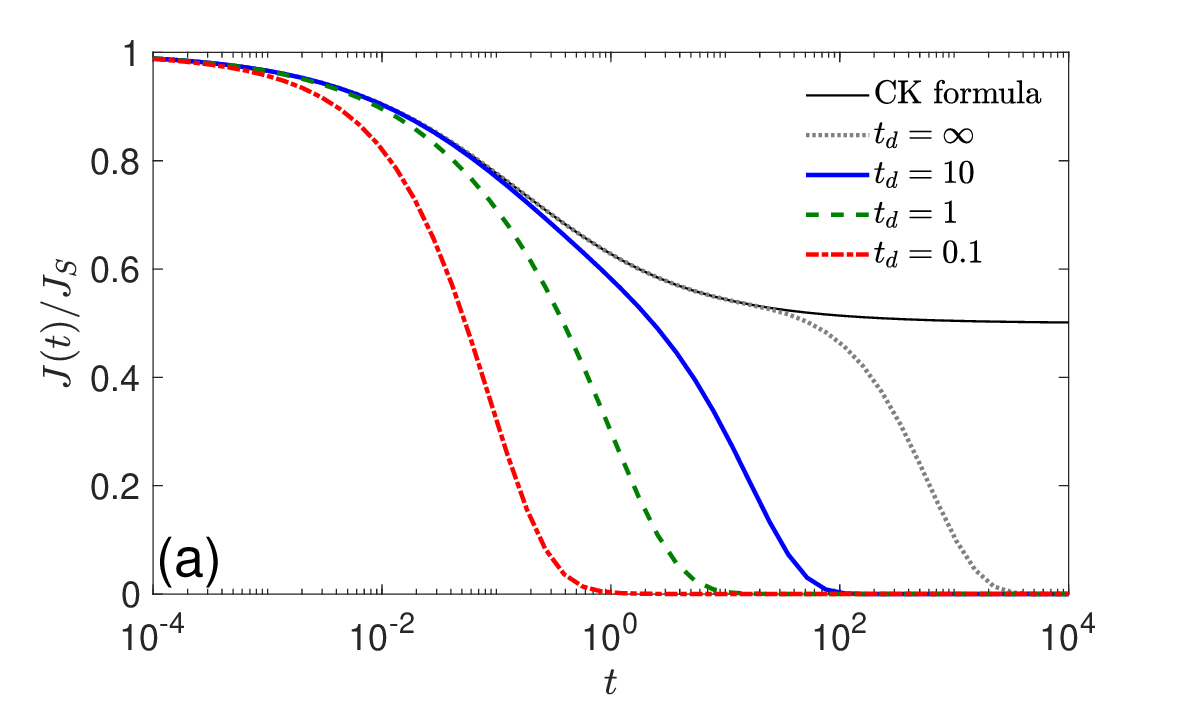} 
\includegraphics[width=85mm]{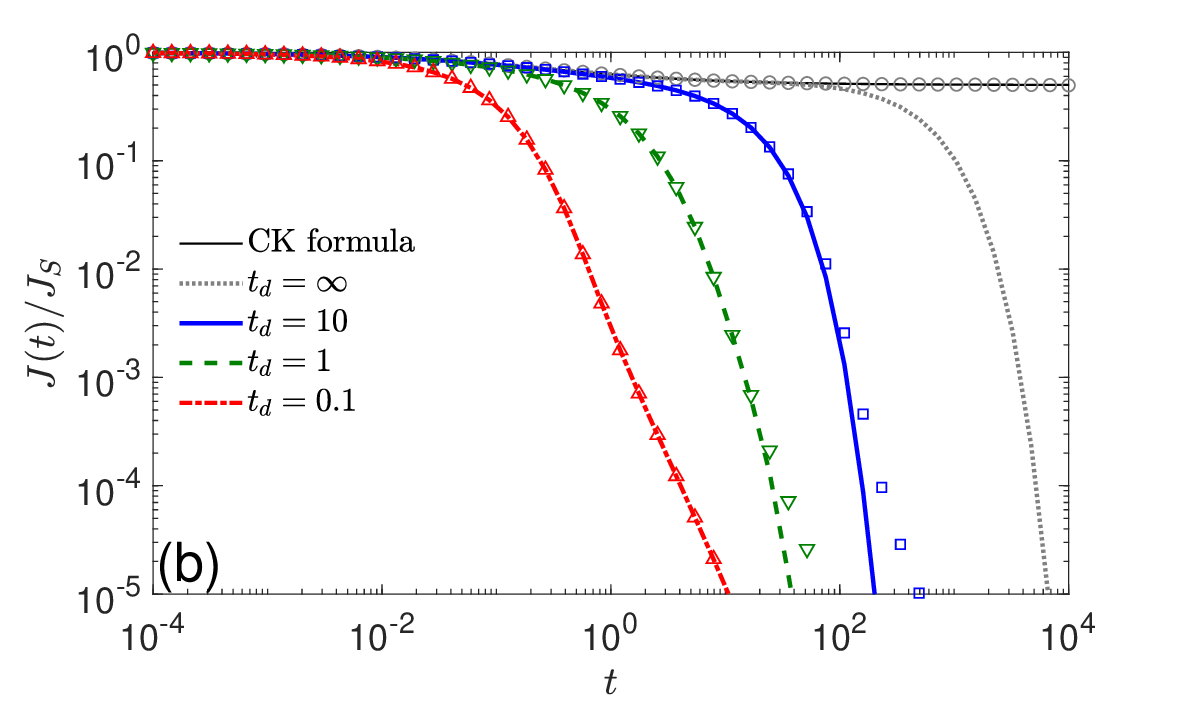} 
\end{center}
\caption{
{\bf (a)} The diffusive flux $J(t)$ divided by the Smoluchowski
steady-state flux $J_{\rm S}$ from Eq. (\ref{eq:Jinf}), for diffusion
between concentric spheres, with $D = 1$, $R = 1$, $L = 10$, and $c_0
= 1$.  Conventional Markovian binding/unbinding kinetics ($\alpha =
\beta = 1$), with $\ell_0 = 1$ (and thus $q = 1$) and four values of
the mean waiting time $\ddelta$ as indicated in the legend.  The
classical Collins-Kimball formula (\ref{eq:Jt_CK}) is compared to the
inverse Laplace transform of the exact expression (\ref{eq:Jp_sphere})
obtained numerically by the Talbot algorithm.  {\bf (b)} The same
diffusive fluxes are presented on the log-log scale, for $L = 10$
(shown by lines, as on panel {\bf (a)}) and $L = 100$ (shown by
symbols).}
\label{fig:Jt}
\end{figure}

Figure \ref{fig:Jt}(a) shows the diffusive flux $J(t)$ as a function
of time $t$ for four values of the mean waiting time $\ddelta$.  Let
us first look at the particular case $\ddelta = \infty$ corresponding
to irreversible binding.  Comparing this curve with the explicit
Collins-Kimball formula (\ref{eq:Jt_CK}) for $L =\infty$ (shown by
thin black line), one can clearly see the effect of confinement.  At
short times, two curves are almost indistinguishable because the flux
is formed by particles near the reactive region and thus there is no
influence of the outer reflecting boundary.  However, as time goes on,
two curves split and start to behave differently.  Indeed, when there
is no outer boundary, the confining domain is unbounded, and there is
an infinite, inexhaustible amount of particles that diffuse toward the
sphere to react on it.  The flux reaches a strictly positive
steady-state limit $J_{\rm S} \frac{qR}{1+qR}$, which depends on the
reactivity parameter $q$.  In contrast, if the domain is bounded, the
amount of particles is limited, and they all irreversibly bind to the
sphere so that the flux drops to $0$ as $t\to\infty$.

The presence of unbinding events (i.e., a finite value of $\ddelta$)
does not fundamentally change the behavior but shifts the curves to
shorter times as $\ddelta$ decreases.  At first thought, this may
sound counter-intuitive because unbinding events ensure that some
particles are present in the confining domain; in particular, we saw
earlier that the fraction of particles in the bulk reaches a constant.
In this case, vanishing of the flux simply reflects that the
equilibrium is reached at long times.

The remarkable effect of unbinding events is that the diffusive flux
does not almost depend on the size $L$ of the confinement if the size
is large enough.  This is illustrated on Fig. \ref{fig:Jt}(b) which
compares the fluxes in two domains, with $L = 10$ (lines) and $L =
100$ (symbols).  For irreversible binding, an enlargement of the
domain to $L = 100$ extends the validity of the Collins-Kimball
formula to longer times, implying the deviation between two curves for
$L = 10$ and $L = 100$.  In contrast, when binding is reversible, the
curves for $L = 10$ and $L = 100$ are almost indistinguishable.
Actually, to see their difference at long times, we had to show the
vertical axis in panel (b) on logarithmic scale.  From the
mathematical point of view, the similarity between these curves comes
from the fact that the smallest eigenvalue $\mu_0^{(p)}$, given by
Eq. (\ref{eq:mu0_sphere}), is exponentially close to its limiting
value $1/R + \sqrt{p/D}$ when $L$ is large enough and $p$ is not too
small (indeed, $\tanh((L-R)\sqrt{p/D}) = 1 + O(exp)$ when
$(L-R)\sqrt{p/D} \gg 1$).  As a consequence, the effect of confinement
can only be seen at very small $p$ or, equivalently, at very long
times.  From the physical point of view, if $L$ is large enough, the
system starts to equilibrate near the reactive sphere (on a time scale
of $\ddelta$), and then the equilibrated layer extends further into
the bulk due to diffusion.  As a consequence, the initial drop of the
diffusive flux $J(t)$ on the reactive sphere does not depend on the
confinement size.

\begin{figure}[t!]
\begin{center}
\includegraphics[width=85mm]{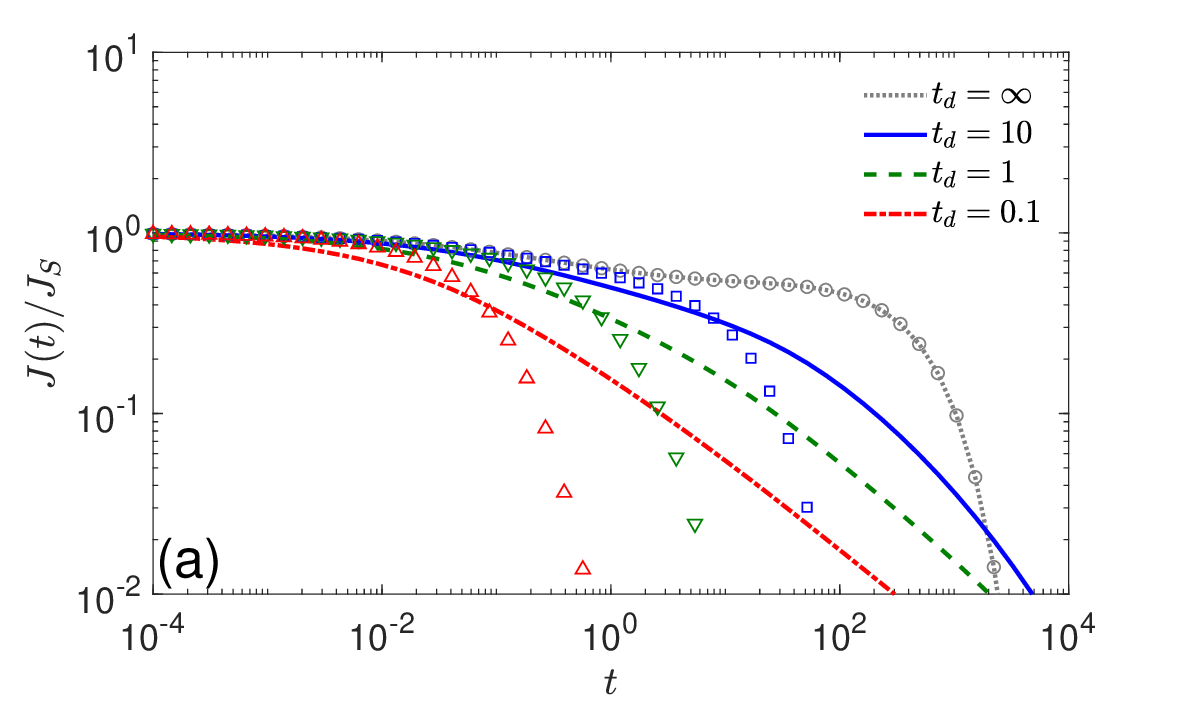} 
\includegraphics[width=85mm]{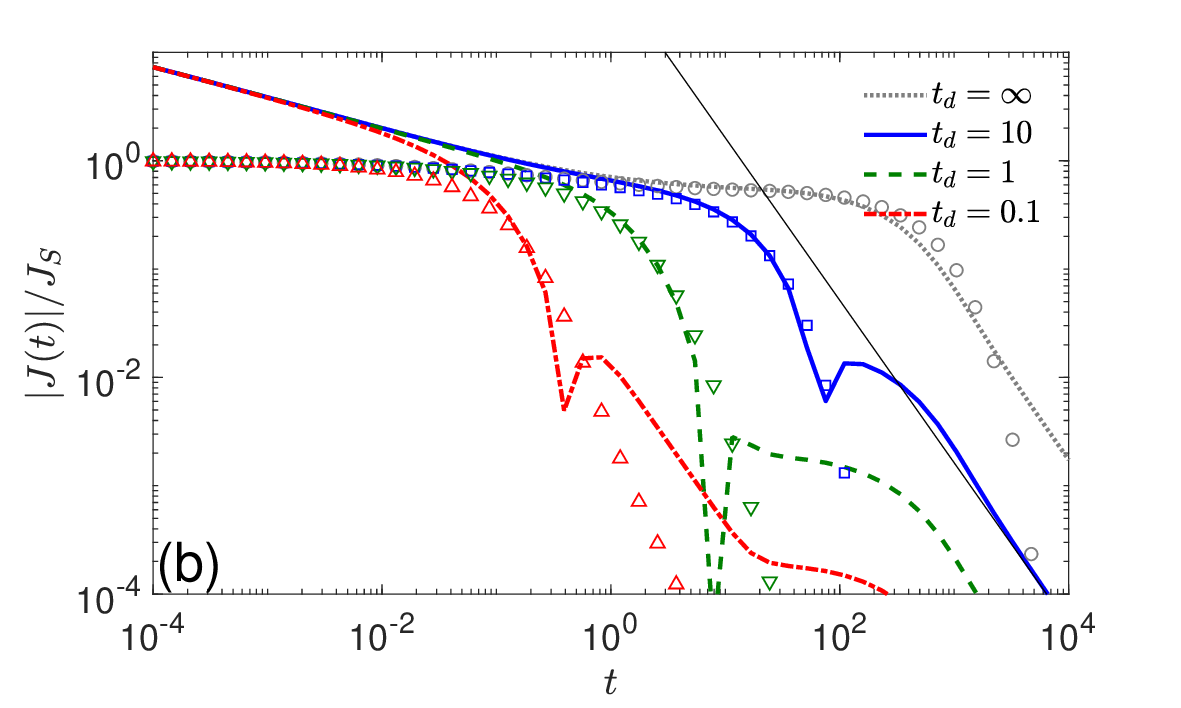} 
\includegraphics[width=85mm]{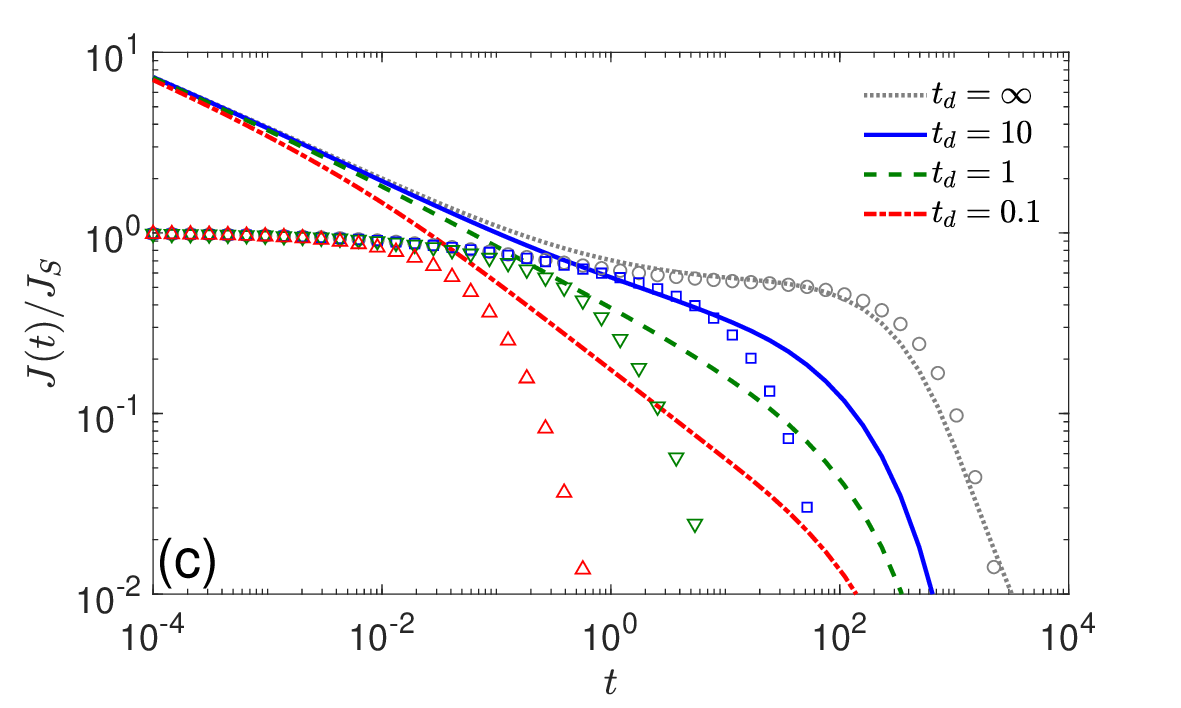} 
\end{center}
\caption{
The diffusive flux $J(t)$ divided by the Smoluchowski steady-state
flux $J_{\rm S}$ from Eq. (\ref{eq:Jinf}), for diffusion between
concentric spheres, with $D = 1$, $R = 1$, $L = 10$, and $c_0 = 1$,
and Mittag-Leffler binding/unbinding kinetics determined by
(\ref{eq:psi_ML}, \ref{eq:phi_ML}), with $\ell_0 = 1$ and four values
of the time scale $\ddelta$ as indicated in the legend.  Symbols
present the reference case of Markovian binding/unbinding kinetics
($\alpha = \beta = 1$). Lines show non-Markovian cases: {\bf (a)}
$\alpha = 1$, $\beta = 0.5$; {\bf (b)} $\alpha = 0.5$, $\beta = 1$;
and {\bf (c)} $\alpha = 0.5$, $\beta = 0.5$.  Black straight line on
panel {\bf (b)} presents the long-time asymptotic relation
(\ref{eq:Jt_sphere_long}) for $\ddelta = 10$. }
\label{fig:Jt2}
\end{figure}

Figure \ref{fig:Jt2} illustrates the effects of non-Markovian binding
and unbinding kinetics.  To highlight these effects, we consider the
aforementioned Markovian kinetics as a reference case (shown by
symbols).  We also fix the threshold scale $\ell_0 = 1$ and use the
same set of time scales $\ddelta$ as in Fig. \ref{fig:Jt}.  Panel (a)
presents the effect of non-Markovian unbinding kinetics ($\alpha =
1,~\beta = 0.5$).  When $\ddelta = \infty$ (irreversible binding), the
unbinding kinetics has no influence, and two curves (for $\beta = 0.5$
and $\beta = 1$) are identical.  In turn, there is a drastic
difference between the cases $\beta = 0.5$ and $\beta = 1$ for any
finite $\ddelta$.  In fact, anomalously long sojourns in the bound
state considerably delay the decay of the diffusive flux and thus the
equilibration of the system.  In line with the long-time asymptotic
analysis of Sec. \ref{sec:long}, one can easily show that the
exponential decay for $\beta = 1$ switches to a power-law decay for
$\beta = 0.5$.  These qualitative conclusions agree with the earlier
results by Agmon and Weiss for the unbounded case $L = \infty$
\cite{Agmon89}.

In turn, panel (b) of Fig. \ref{fig:Jt2} presents the new effect of
non-Markovian binding kinetics, while keeping the unbinding events to
be Markovian ($\alpha = 0.5,~\beta = 1$).  Let us first examine the
short-time behavior.  As the mean threshold is infinite in the case
$\alpha = 0.5$, binding events are expected to be rarer, as compared
to the case of Markovian binding with $\alpha = 1$, and therefore it
would be more difficult for a particle to adsorb on the substrate.
One might thus expect that the diffusive flux would be smaller for
$\alpha = 0.5$ than for $\alpha = 1$.  This is not the case.  In
contrast, while the diffusive flux converged to a constant as $t\to 0$
for the Markovian binding kinetics (in agreement with the
Collins-Kimball formula), it diverges for the considered non-Markovian
setting.  In fact, substituting $\mu_0^{(p)} \approx\sqrt{p/D}$ and
$\dtilde{\psi}(\mu) \approx (\mu \ell_0)^{-\alpha}$ for large $p$ and
$\mu$ into Eq. (\ref{eq:Jtilde_sphere}), one gets
\begin{equation}
\tilde{J}(p) \approx J_{\rm S} \frac{R D^{(\alpha-1)/2}}{\ell_0^\alpha} \, p^{-(\alpha+1)/2} \quad (p\to \infty),
\end{equation}
from which
\begin{equation}
J(t) \approx J_{\rm S} \frac{ R \, (Dt/\ell_0^2)^{(\alpha-1)/2}}{\ell_0 \, \Gamma(\frac{\alpha+1}{2})}  \quad (t\to 0),
\end{equation}
in agreement with the power-law divergence seen in
Fig. \ref{fig:Jt2}(b).  Expectedly, the short-time behavior does not
depend on the unbinding kinetics so that four curves for different
$\ddelta$ fall onto each other.  

What is the reason of this enhanced flux?  We recall that the
encounter-based description identifies the binding event with the
first crossing of the random threshold $\hat{\ell}$ by the boundary
local time.  According to Eq. (\ref{eq:ML_asympt0}), small values of
the threshold $\hat{\ell}$ are more likely than in the Markovian
model.  As a consequence, the particle can easier bind the reactive
region at short times within the Mittag-Leffler model.  Evidently,
this peculiar result is specific to the small-$\ell$ behavior of the
probability density $\psi(\ell)$.  In \cite{Grebenkov20}, other models
were discussed, for which the diffusive flux may tend zero as $t\to
0$.  We finally note that the strongest divergence $J(t) \propto
t^{-1/2}$ corresponds to the formal limit $\alpha = 0$, which
corresponds to the Smoluchowski setting of a perfect adsorption, see
Eq. (\ref{eq:Jt_Smol}).

The long-time behavior of the diffusive flux on Fig. \ref{fig:Jt2}(b)
reveals another interesting feature.  At intermediate times, the
curves for $\alpha = 0.5$ are close to those for $\alpha = 1$
suggesting that the distinction between Markovian and non-Markovian
binding kinetics is progressively reduced due to several
binding/unbinding events.  However, at longer times, the curves
separate again and become drastically different.  In particular, one
can notice a spike-like feature which actually means that the
diffusive flux becomes negative (note also that the absolute value of
the diffusive flux is plotted on this panel, to be able to employ the
logarithmic scale on the vertical axis).  The negative values of the
diffusive flux can be also deduced from the long-time asymptotic
analysis from Sec. \ref{sec:long}.  Repeating it for the diffusive
flux, one gets
\begin{equation}
\tilde{J}(p) \approx J_{\rm S}  \frac{R \ttau^{1-\alpha} \ddelta}{\ell_0} \, p^{1-\alpha}  \quad (p \to 0),
\end{equation}
so that
\begin{equation}  \label{eq:Jt_sphere_long}
J(t) \approx J_{\rm S}  \frac{R \ttau^{1-\alpha} \ddelta}{\ell_0 \Gamma(\alpha-1)} \, t^{\alpha-2}  \quad (t \to \infty).
\end{equation}
Since $0< \alpha < 1$, $\Gamma(\alpha-1)$ is negative, and so is the
flux at long times.  This behavior can be intuitively expected.  At
the beginning, there is no particle in the bound state, and eventual
binding events are responsible for the positive flux at short and
intermediate times.  However, the power-law decay in
Eq. (\ref{eq:ML_asymptinf}) allows for anomalously large values of the
random threshold $\hat{\ell}$ so that binding events are getting rarer
and rarer in the course of time.  As a consequence, the bound
particles start to be released more often than the new ones are
getting bound, implying the negative flux.  The power-law decay
(\ref{eq:Jt_sphere_long}) implies that the equilibration of the system
is anomalously long.  We stress that the negative flux can be also
found in the Markovian setting; it is actually related to the
non-monotonous behavior of the probability $S(t|\x_0)$ (see, e.g., the
curve shown by blue circles on Fig. \ref{fig:Qt}(b)).

The last panel (c) of Fig. \ref{fig:Jt2} illustrates the combined
effect of both non-Markovian binding and unbinding kinetics ($\alpha =
\beta = 0.5$).  Expectedly, the short-time behavior has not changed as
being unaffected by unbinding kinetics.  In turn, the long-time
behavior is again modified.  In this particular example, the diffusive
flux remains positive for all times but still exhibits a slow
power-law decay.  Its positive character simply reflects a sort of
balance between anomalously rare binding events and anomalously long
durations of unbinding events.

\subsection{Concentration evolution}

\begin{figure}[t!]
\begin{center}
\includegraphics[width=42mm]{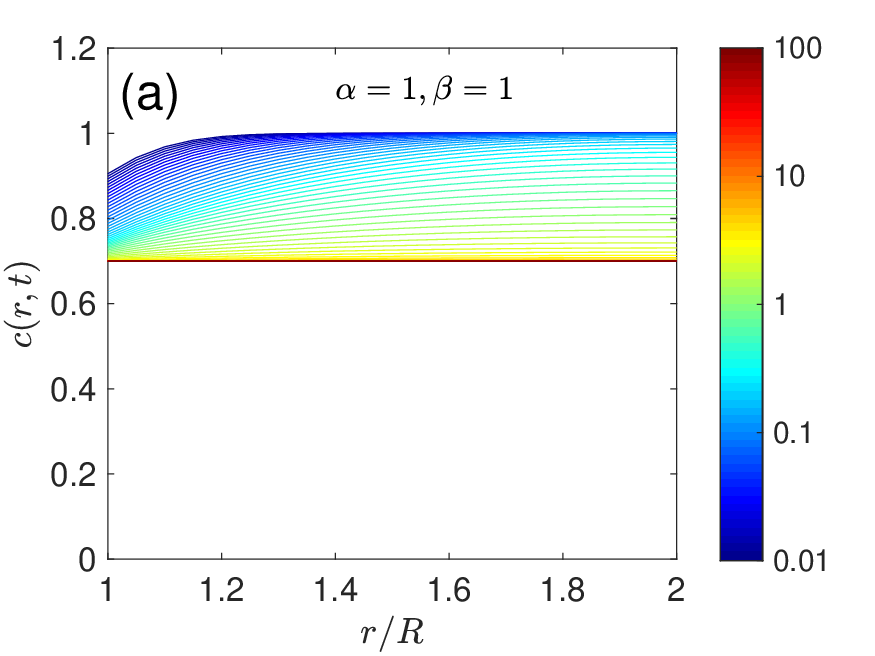} 
\includegraphics[width=42mm]{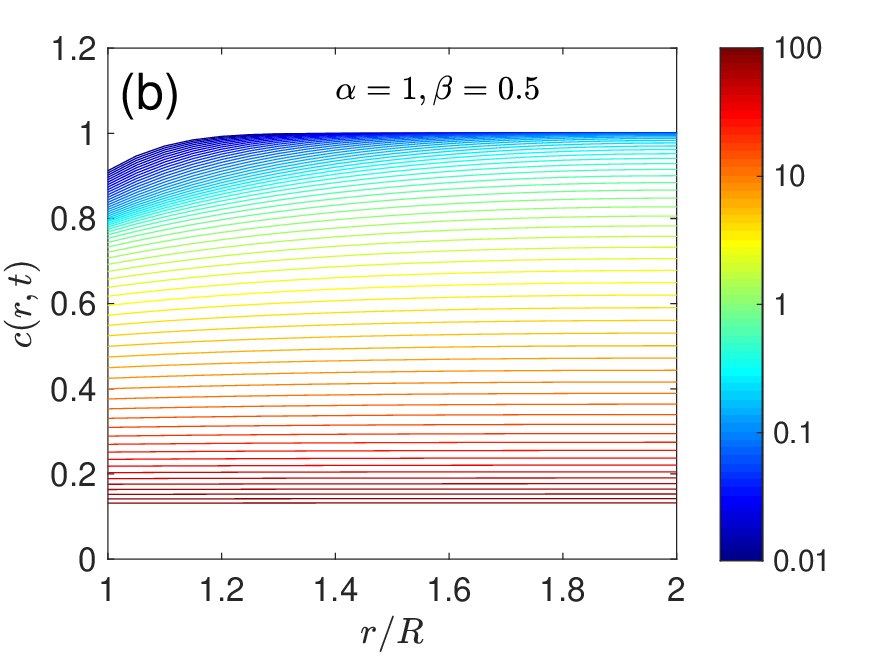} 
\includegraphics[width=42mm]{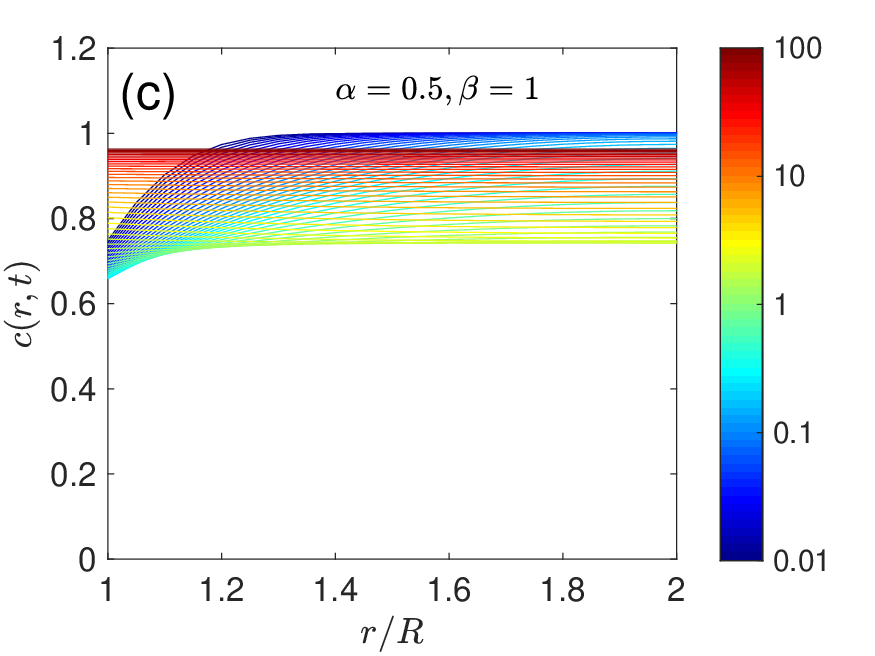} 
\includegraphics[width=42mm]{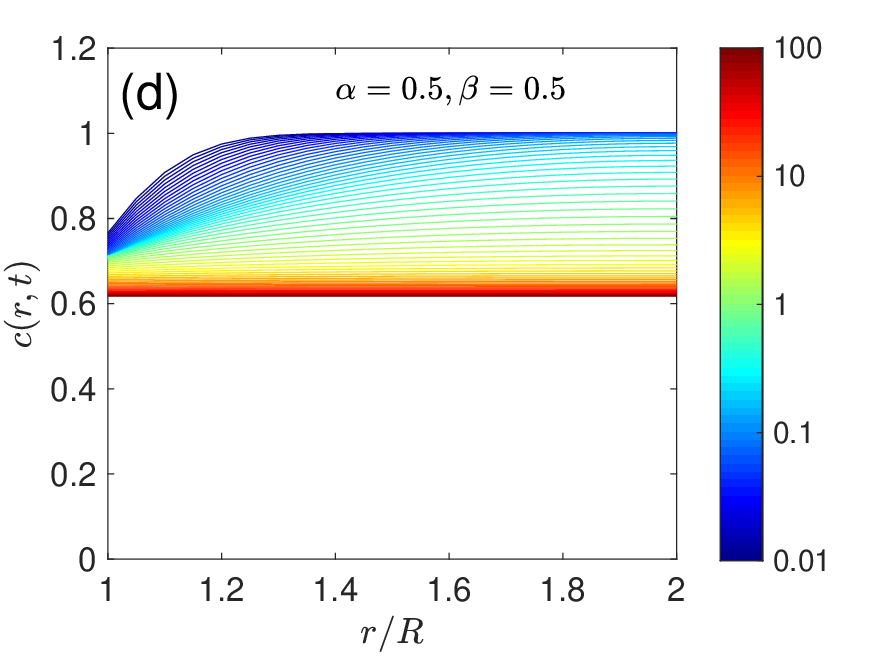} 
\end{center}
\caption{
The concentration profile $c(\x,t)$ (rescaled by $c_0$) as a function
of $r = |\x|$, for diffusion between concentric spheres, with $D = 1$,
$R = 1$, $L = 2$, $c_0 = 1$, and the Mittag-Leffler binding/unbinding
kinetics determined by (\ref{eq:psi_ML}, \ref{eq:phi_ML}), with
$\ell_0 = 1$ and $\ddelta = 1$.  The concentration was evaluated at 64
times ranging uniformly on the logarithmic scale from $0.01$ to $100$
and represented by colors, from dark blue ($t = 0.01$) to dark red ($t
= 100$).  {\bf (a)} $\alpha = 1$, $\beta = 1$; {\bf (b)} $\alpha = 1$,
$\beta = 0.5$; {\bf (c)} $\alpha = 0.5$, $\beta = 1$; and {\bf (d)}
$\alpha = 0.5$, $\beta = 0.5$.  }
\label{fig:cxt}
\end{figure}

Yet another insight onto reversible diffusion-controlled reactions can
be achieved by looking at the temporal evolution of the concentration
profile $c(\x,t)$ from the uniform initial concentration $c_0$.  We
recall that this quantity is proportional to the probability $S(t|\x)$
and is a function of the radial distance $r = |\x|$ for the considered
spherical domain.  Figure \ref{fig:cxt} illustrates how the
concentration evolves with time for four combinations of the exponents
$\alpha$ and $\beta$ (all other parameters being kept fixed to ease
the comparison).  In the conventional setting $\alpha = \beta = 1$
(panel (a)), one observes an exponential approach to the equilibrium
concentration, $c_0/(1 + \ddelta/\ttau) = 0.7$, as discussed in
Sec. \ref{sec:long}.  When the unbinding kinetics is non-Markovian
(panel (b)), the particles stay longer and longer in the bound state
so that the concentration of unbound particles slowly vanishes.  In
turn, if the binding kinetics is non-Markovian (panel (c)), binding
events are rare at long times, and the concentration is slowly
restored to the initial level $c_0$.  Finally, when both kinetics are
non-Markovian (panel (d)), the rarity of binding and unbinding events
is compensated and leads to a slow relaxation to a new equilibrium
concentration, $c_0/(1 + (\ddelta/\ttau)^\alpha) \approx 0.60$.

\section{Discussion}
\label{sec:discussion}

Despite successful applications of the conventional (Markovian)
binding/unbinding kinetics, its numerous limitations have been
identified long ago.  For instance, Agmon and Weiss introduced a
general waiting time distribution to account for anomalously long
sojourns in the bound states \cite{Agmon89}.  Similarly, the simple
form of the forward reaction term in Eq. (\ref{eq:Robin_rev1}) cannot
fully account for microscopic heterogeneity of the reactive region,
its temporal variations, saturation effects in adsorption processes
and passivation of catalysts, as well as various regulation and
control mechanisms of biochemical reactions in living cells.  Quite
naturally, different extensions of the forward term have been
proposed.  Here we briefly discuss a large class of extensions that
impact reactivity but do not alter the linearity of the forward term
with respect to $c(\x,t)$.

There are at least three basic ways to render reactivity nonconstant
(and thus to go beyond the classical setting): (i) a space-dependent
reactivity $\kappa(\x)$ that allows one to model microscopic
heterogeneity of the reactive region; (ii) a time-dependent reactivity
$\kappa(t)$ that implements a temporal evolution of that region; and
(iii) a $p$-dependent reactivity $\tilde{\kappa}(p)$ in the Laplace
domain that incorporates memory effects via a convolution-type
boundary condition (\ref{eq:Robin_conv}) in time domain (one can also
consider their combinations).  The first extension is in general
challenging because the diffusive dynamics is intrinsicly coupled here
to surface reactions: the particle encounters the reactive region in
different locations and thus realizes a sequence of reaction attempts
with different reaction probabilities that depend on the whole random
trajectory of the particle.  A general spectral approach for this
extension was proposed in \cite{Grebenkov19b} (see further discussions
and references therein).  The second extension is also difficult for
analytical treatments because the Laplace transform, which is usually
performed to reduce the diffusion equation to the modified Helmholtz
equation, would lead to a convolution-type Robin boundary condition in
the Laplace domain.  In contrast, the third extension, that we
discussed in Sec. \ref{sec:bindingM}, does not change the solution in
the Laplace domain, except that the constant reactivity $\kappa_0$ is
replaced by the reactivity $\tilde{\kappa}(p)$ that depends on $p$ as
a fixed parameter.  In turn, the asymptotic behavior of the solution,
as well as its form in time domain, are modified.

The general binding mechanism discussed in this paper can be
associated to the fourth extension of a constant reactivity.  As first
suggested in \cite{Grebenkov20}, the probability density $\psi(\ell)$
of the threshold $\hat{\ell}$ can be related to the so-called {\it
encounter-dependent reactivity} as
\begin{equation}  \label{eq:kappa_psi}
\kappa(\ell) = D \frac{\psi(\ell)}{\int\nolimits_{\ell}^\infty d\ell' \, \psi(\ell')} \,.
\end{equation}
In the conventional setting, the particle attempts to react at each
arrival onto $\pa_R$ with the constant probability $\rho \approx a
\kappa_0/D$ (see Sec. \ref{sec:encounter}).  In turn,
Eq. (\ref{eq:kappa_psi}) allows one to implement an arbitrary
dependence of the reaction probability on the number of encounters,
i.e., on the number of failed attempts represented by $\ell$.  The
encounter-dependent reactivity can be interpreted in a similar way to
the time-dependent diffusivity $D(t)$ \cite{Latour94,Sen04,Wu08}.  In
fact, if the bulk properties change with time, their effect onto
diffusive displacements can be modeled via a prescribed dependence
$D(t)$.  Likewise, if the reactive properties of the substrate depend
on the number of encounters with the particle, it can be modeled via a
prescribed dependence $\kappa(\ell)$.  This is yet another
manifestation of the fundamental similarity between the physical time
as a proxy of the number of bulk jumps and the boundary local time as
a proxy of the number of jumps onto the boundary.

To illustrate the concept of the encounter-dependent reactivity, let
us inspect again the Mittag-Leffler model.  Inserting
Eq. (\ref{eq:psi_ML}) for the probability density $\psi(\ell)$ in
Eq. (\ref{eq:kappa_psi}), one gets
\begin{equation}  \label{eq:kappa_ML}
\kappa(\ell) = \kappa_0 (\ell/\ell_0)^{\alpha-1} 
\frac{E_{\alpha,\alpha}(-(\ell/\ell_0)^\alpha)}{E_{\alpha,1}(-(\ell/\ell_0)^\alpha)} \,,
\end{equation}
with $\kappa_0 = D/\ell_0$.  This function is shown on
Fig. \ref{fig:psi}(b).  In the Markovian case $\alpha = 1$, one simply
retrieves a constant reactivity $\kappa(\ell) = \kappa_0$, which
stands in the Robin boundary condition (\ref{eq:Robin}).  In turn, if
$0 < \alpha < 1$, the above encounter-dependent reactivity exhibits
two asymptotic power-law behaviors
\begin{subequations}  \label{eq:kappa_ML_asympt}
\begin{align}
\kappa(\ell) & \approx \frac{\kappa_0}{\Gamma(\alpha)} (\ell/\ell_0)^{\alpha-1}  \quad (\ell \to 0)  ,\\
\kappa(\ell) & \approx \kappa_0 \alpha (\ell/\ell_0)^{-1}  \quad (\ell \to \infty)  .
\end{align}
\end{subequations}
In the limit $\ell\to 0$, the encounter-dependent reactivity diverges,
indicating that $\pa_R$ is highly reactive.  This is consistent with
the earlier discussion that small values of the threshold $\hat{\ell}$
are more likely to occur than for the Markovian setting with $\alpha =
1$.  The higher reactivity can thus explain the larger diffusive flux
observed in Fig. \ref{fig:Jt2}(b).  In the opposite limit $\ell\to
\infty$, the reactivity slowly decays, resulting in rarer and rarer
binding events.  While we focused on the Mittag-Leffler model for
illustrative purposes, one can easily implement any function
$\kappa(\ell)$, which is suitable to represent the reactivity
evolution of the substrate, by using the random threshold with the
probability density
\begin{equation}
\psi(\ell) = \frac{\kappa(\ell)}{D} \exp\left(-\frac{1}{D} \int\limits_0^\ell d\ell' \, \kappa(\ell')\right).
\end{equation}
In other words, there is a mapping between $\kappa(\ell)$ and
$\psi(\ell)$.  In \cite{Grebenkov20}, several other models and their
consequences for irreversible surface reactions were discussed.

We emphasize that $\kappa(\ell)$ depends on the number of encounters
(represented by $\ell$), not on the physical time $t$.  In other
words, we model here the dynamical situation when the reactivity of
the target $\pa_R$ changes due to its ``interactions'' with the
particle, and not because of some external actions.  We also stress
that the effect of the encounter-dependent reactivity $\kappa(\ell)$
onto diffusion-controlled reactions is not in general reduced to a
$p$-dependent reactivity $\tilde{\kappa}(p)$ or to a convolution-type
boundary condition (\ref{eq:Robin_conv}) with a memory kernel $\K(t)$,
discussed in \ref{sec:bindingM}.  This reduction is only possible for
some quantities and some specific geometric settings such as the
spherical target considered in Sec. \ref{sec:sphere}.

In order to improve the intuitive comprehension of the general binding
mechanism, let us further highlight its similarity with the
description of unbinding events.  In fact, the long history of
continuous-time random walks gets us used to the concept of waiting
times with heavy-tailed distributions and the consequent anomalous
features \cite{Montroll65,Metzler00,Klafter}.  One can formally say
that an unbinding event occurs at the first instance $\tau$ when the
time spent by the particle in the bound state exceeds a random
threshold $\hat{\tau}$ obeying a given probability density $\phi(t)$:
\begin{equation}
\tau = \inf\{ t > 0 ~:~ t > \hat{\tau}\} .
\end{equation}
However, this is a tautology because $\tau = \hat{\tau}$.  At the same
time, this formal definition provides a direct analogy to the
introduction of the first instance of a binding event via
Eq. (\ref{eq:Tdef}).  As the boundary local time $\ell_t$ is the proxy
of the number of encounters between the particle and the reactive
region, the sequence of failed reaction attempts is stopped when
$\ell_t$ exceeds an appropriate random threshold $\hat{\ell}$ with a
given probability density $\psi(\ell)$.  In this light, both binding
and unbinding events are implemented in the same way, with the only
difference that the ``counter'' of failed unbinding attempts is the
physical time $t$, whereas the ``counter'' of failed binding attempts
is the boundary local time $\ell_t$.  While the impact of the waiting
time and thus of a ``threshold'' $\hat{\tau}$ onto diffusive processes
had been studied for many decades, the very similar threshold
$\hat{\ell}$ for binding events, which was introduced in
\cite{Grebenkov06,Grebenkov07,Grebenkov19a,Grebenkov20}, has not yet
got a proper attention.

\section{Conclusion}
\label{sec:conclusion}

In this paper, we developed a general theory of reversible
diffusion-controlled reactions with general, non-Markovian binding and
unbinding kinetics.  In this theory, surface reactions are described
by two functions: the probability density $\phi(t)$ that characterizes
the random waiting time of a particle in the bound state, and the
probability density $\psi(\ell)$ of the random number of failed
reaction attempts prior to the successful binding.  While an extension
of the classical theory of reversible reactions to unbinding events
with a general waiting time distribution was proposed by Agmon and
Weiss more than thirty years ago \cite{Agmon89}, an implementation of
non-Markovian binding events required recent advances on the
encounter-based approach \cite{Grebenkov20}.  Combining this approach
with the renewal technique, we managed to derive the spectral
expansion (\ref{eq:Qtilde}) for the Laplace-transformed propagator
$\tilde{Q}(\x,p|\x_0)$.  When both $\phi(t)$ and $\psi(\ell)$ are
exponential, one retrieves the Markovian setting with conventional
forward rate constant $\kon$ (or $\kappa_0$) and backward rate
$\koff$.  From the propagator, we deduced other quantities such as the
concentration of particles and the diffusive flux onto the reactive
region.
While spectral expansions are very common for describing
diffusion-controlled reactions in bounded domains, they are usually
based on the eigenmodes of the Laplace operator (or the Fokker-Planck
operator) \cite{Redner,Gardiner,VanKampen}.  In turn,
Eq. (\ref{eq:Qtilde}) employs the eigenmodes of the
Dirichlet-to-Neumann operator $\M_p$, which is most suitable for
describing diffusive exploration of the bulk between consecutive
binding events.  Being less known than the Laplace operator, the
operator $\M_p$ offers flexible complementary tools for studying
diffusion-controlled reactions.

As most formulas were derived in the Laplace domain, a numerical
inversion of the Laplace transform was needed to present the results
in time domain.  Even without this inversion, the Laplace-transformed
quantities allow one to determine the asymptotic behavior at short and
long times.  Moreover, if the diffusing particle has a random lifetime
$\delta$ (so-called ``mortal'' random walker
\cite{Yuste13,Meerson15,Grebenkov17f,Meerson19}), the
Laplace-transformed quantities admit additional probabilistic
interpretations.  For instance,
\begin{equation}
p \tilde{Q}(\x,p|\x_0) = \int\limits_0^\infty dt \, p e^{-pt} \, Q(\x,t|\x_0) = \E\{ Q(\x,\delta|\x_0)\} 
\end{equation}
is the probability density of finding the particle in a vicinity of a
bulk point $\x$ at the death time $\delta$, where $pe^{-pt}$ is the
probability density of $\delta$ with the rate $p$.  In other words,
one does not need to perform the inverse Laplace transform in the
framework of mortal random walkers.

When the binding kinetics is Markovian, the effect of non-Markovian
unbinding events is captured in the Laplace domain via a $p$-dependent
reactivity $\tilde{\kappa}(p)$, which is related by
Eq. (\ref{eq:qp_def}) to the probability density $\phi(t)$.  In time
domain, one retrieves thus a convolution-type boundary condition
(\ref{eq:Robin_conv}) with a memory kernel $\K(t) =
\kappa_0\bigl(\delta(t) - \phi(t)\bigr)$.  This result was already
reported by Agmon and Weiss for the specific case of a spherical
target \cite{Agmon89}.  Our approach permitted to generalize it to
arbitrary bounded domains.  In turn, when the binding kinetics is
non-Markovian, the natural description of binding events involves the
encounter-dependent reactivity $\kappa(\ell)$.  In general, its
effects are not reducible to an effective $\tilde{\kappa}(p)$ or to a
memory kernel $\K(t)$.  Further exploration of its effects and
potentials for describing realistic surface reactions presents an
important perspective for the future.

We also investigated the long-time behavior of the propagator
$Q(\x,t|\x_0)$ and related quantities.  Depending on the exponents
$\alpha$ and $\beta$ that characterize the asymptotic behavior of the
densities $\psi(\ell)$ and $\phi(t)$, four regimes were distinguished.
When $\alpha = \beta = 1$, one retrieves the classical exponential
relaxation toward the uniform equilibrium state.  If $\alpha <
\beta$, binding events occur rarer than unbinding ones at long times
so that the particle will be asymptotically in the unbound state; in
turn, the opposite inequality $\alpha > \beta$ makes binding events
more frequent so that the particle will be mostly in the bound state.
Finally, if $\alpha = \beta < 1$, a subtle balance between anomalously
rare binding and unbinding events is settled, and a new uniform
equilibrium state is approached anomalously slowly.

In order to highlight the effects of non-Markovian binding/unbinding
kinetics, we considered the most basic setting of a single particle
undergoing ordinary diffusion in the bulk.  This study can be extended
in several directions.  (i) The ordinary bulk diffusion can be
replaced by more general stochastic processes such as continuous-time
random walks \cite{Montroll65,Metzler00,Klafter}, processes with
diffusing diffusivity \cite{Chubynsky14,Chechkin17,Lanoiselee18},
diffusion with a drift or in an external potential \cite{Grebenkov22a}
diffusion with stochastic resetting
\cite{Evans20,Bressloff22d,Benkhadaj22}, or in the presence of
escape regions \cite{Grebenkov23}.  (ii) The collective effect of
multiple independently diffusing particles can be implemented by
defining binding events through the total boundary local time that
these particles spent on the reactive region
\cite{Grebenkov22b}.  (iii) When the reactive region is bounded,
the encounter-based approach can be used even for unbounded domains
(see \cite{Grebenkov21}); however, as briefly discussed in
Sec. \ref{sec:sphereInf}, the long-time behavior can be considerably
altered.  (iv) In the case of a small reactive region, one can employ
matched asymptotic techniques and other approximations
\cite{Bressloff22b,Grebenkov22c} to capture more explicitly the
effects of non-Markovian binding/unbinding kinetics.  (v) The majority
of former contributions to reversible diffusion-controlled reactions
dealt either with one-dimensional setting (diffusion on a half-line
or, equivalently, diffusion in the half-space), or with diffusion
outside a reactive sphere.  The symmetry of these domains helped to
implement adsorption/desorption kinetics rather explicitly (e.g., see
Sec. \ref{sec:sphere}), but did not allow one to fully explore new
features related to the encounter-dependent reactivity.  In the
future, it would be interesting to investigate how non-Markovian
binding kinetics may affect diffusion-controlled reactions in more
general confining domains.

\section*{Data Availability Statement}

Data sharing is not applicable to this article as no new data were
created or analyzed in this study.

\begin{acknowledgments}
The author acknowledges the Alexander von Humboldt Foundation for
support within a Bessel Prize award.
\end{acknowledgments}

\appendix
\section{A microscopic derivation of Robin boundary conditions}
\label{sec:micro}

In this Appendix, we provide a simple derivation of Robin-type
boundary conditions for reversible binding in the conventional case of
Markovian binding and unbinding kinetics.  While the origins of Robin
boundary condition and its microscopic interpretations for
irreversible reactions have been thoroughly discussed
\cite{Berg77,Szabo89,Grebenkov03,Grebenkov07,Singer08,Lawley15,Chapman16,Grebenkov19h,Piazza22},
we propose here a simple derivation that may be instructive for
non-expert readers.

\begin{figure}[t!]
\begin{center}
\includegraphics[width=45mm]{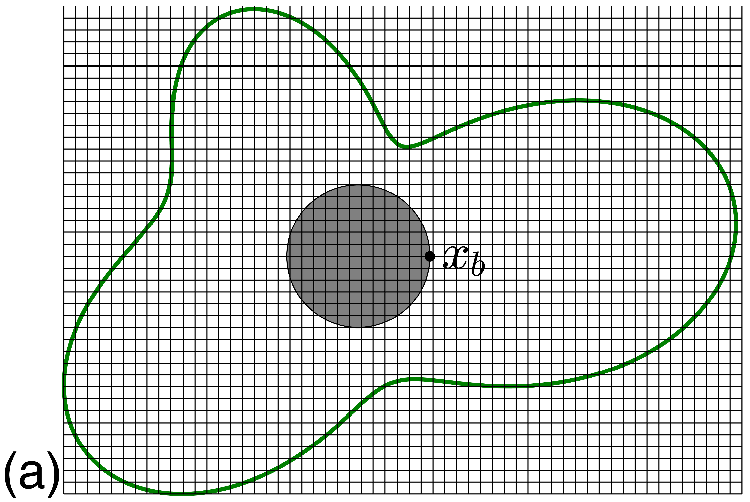} 
\includegraphics[width=40mm]{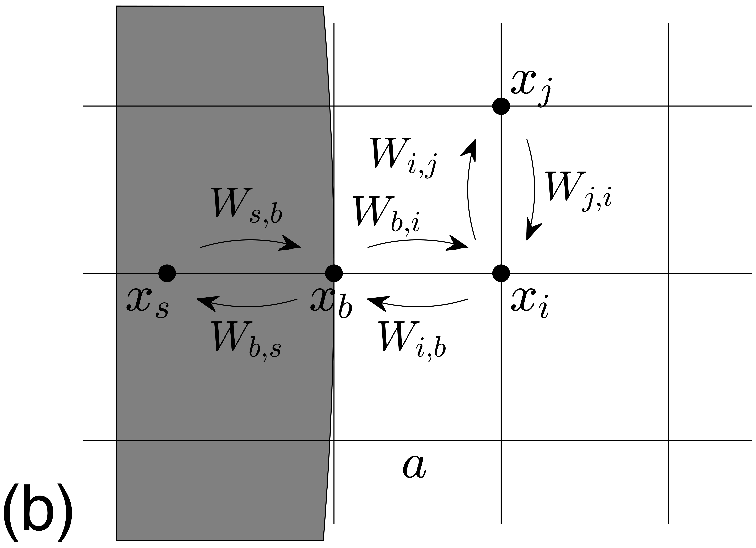} 
\end{center}
\caption{
{\bf (a)} A confining domain $\Omega\subset \R^d$ in two dimensions
($d = 2$) is discretized by a square lattice.  {\bf (b)} A zoom of the
vicinity of a boundary point $\x_b$, which is connected to a single
bulk point $\x_i$ in the interior and to a storage site $\x_s$.  Some
transition rates $W$ between connected sites are shown.}
\label{fig:domain_grid}
\end{figure}

To get an intuitive picture of Robin boundary conditions, let us
approximate Brownian motion by a symmetric random walk on a
$d$-dimensional (hyper)cubic lattice with a step $a$ that discretizes
the confining domain $\Omega$ (Fig. \ref{fig:domain_grid}(a)).  As
surface reactions are local in space, it is sufficient to consider a
vicinity of a boundary point $\x_b$ on $\pa_R$
(Fig. \ref{fig:domain_grid}(b)).  For the sake of simplicity, we
assume that this point is connected to a single bulk point $\x_i$ in
the interior of the domain (at distance $a$ from $\x_b$).  As the
binding event consists in a temporal ``storage'' of the particle on
the boundary, the surface point $\x_b$ is also connected to a storage
site $\x_s$ behind it.  The diffusive dynamics of this random walk can
be described by the master equation for the probability $q(\x_k,t)$ of
finding the particle in a state $\x_k$ at time $t$.  For three
aforementioned sites, the master equation reads
\begin{subequations}
\begin{align}  \label{eq:master1}
\frac{dq(\x_s,t)}{dt} & = - q(\x_s, t) W_{s,b} + q(\x_b,t) W_{b,s}  ,\\  \nonumber
\frac{dq(\x_b,t)}{dt} & = - q(\x_b, t) \bigl[W_{b,s} + W_{b,i}\bigr] \\   \label{eq:master2}
& + q(\x_i,t) W_{i,b} + q(\x_s,t) W_{s,b}  ,\\  \label{eq:master3}
\frac{dq(\x_i,t)}{dt} & = - q(\x_i, t) \sum_j W_{i,j} + \sum_j q(\x_j,t) W_{j,i}   ,
\end{align}
\end{subequations}
where the sums in the last relation are taken over all $2d$ neighbors
$\x_j$ of the bulk site $\x_i$, and $W_{i,j}$ denotes the transition
rate from site $i$ to site $j$.  For bulk diffusion, one has
\begin{equation}
W_{j,i} = W_{i,j} = \frac{1}{2d\delta} \,,
\end{equation}
where $\delta$ is a time step of one jump.  The last master equation
(\ref{eq:master3}) is the discrete version of the diffusion equation
with
\begin{equation}  \label{eq:D_a}
D = \frac{a^2}{2d\delta} \,.  
\end{equation}
Using this relation, the second master
equation (\ref{eq:master2}) can be written as
\begin{align*}
\frac{dq(\x_b,t)}{dt} & = - \frac{D}{a} \bigl(\partial_n q(\x,t)\bigr)_{\x=\x_b} \\
& - q(\x_b,t) W_{b,s} + q(\x_s,t) W_{s,b}  ,
\end{align*}
where we replaced the discretized form of the normal derivative,
$(q(\x_i,t) - q(\x_b,t))/a$, by its continuous form.  Substituting
$W_{s,b} = \koff$ and dividing by $a^{d-1}$, we get
\begin{align} \nonumber
a \frac{dc(\x_b,t)}{dt} & = - D \bigl(\partial_n c(\x,t)\bigr)_{\x=\x_b} \\    \label{eq:master2b}
& - \kappa_0 c(\x_b,t)  + \koff c_{b}(\x_b, t) ,
\end{align}
where we set 
\begin{equation}  \label{eq:kappa_W}
\kappa_0 = a W_{b,s}  
\end{equation}
and introduced the bulk and surface concentrations as
\begin{equation}
c(\x,t) = N_0 \frac{q(\x,t)}{a^d},  \quad c_b(\x_b,t) = N_0 \frac{q(\x_s,t)}{a^{d-1}} .
\end{equation}
Here we included the number $N_0$ of independent diffusing particles
to pass from a single-molecule description to the macroscopic one
(expectedly, the surface concentration has the units 1/m$^{d-1}$ or
mol/m$^{d-1}$).  Note also that the storage site $\x_s$ was identified
with the associated boundary site $\x_b$.  In the limit $a \to 0$, the
left-hand side of Eq. (\ref{eq:master2b}) vanishes, and one gets the
boundary condition
\begin{equation}  \label{eq:master2c}
- D \partial_n c(\x,t) = \kappa_0 \, c(\x,t) - \koff \, c_{b}(\x,t)   \quad (\x\in\pa_R).
\end{equation}
Finally, the division of the first master equation (\ref{eq:master1})
by $a^{d-1}$ yields
\begin{equation}  \label{eq:master1c}
\frac{dc_b(\x_b,t)}{dt} = - \koff \, c_b(\x_b, t) + \kappa_0 \, c(\x_b,t) ,
\end{equation}
which is identical to Eq. (\ref{eq:Robin_rev1}).  In turn, the
combination of Eqs. (\ref{eq:master2c}, \ref{eq:master1c}) yields
Eq. (\ref{eq:Robin_rev2}).

When there is no unbinding event (i.e., $\koff = 0$), the particle
stays in the bound state forever, and Eq. (\ref{eq:master2c}) is
reduced to the conventional Robin boundary condition.  Here, the
left-hand side is the diffusive flux from the bulk to the boundary
(i.e., from $\x_i$ to $\x_b$ in the discrete picture), whereas the
right-hand side is the ``reactive flux'' into the bound state (i.e.,
from $\x_b$ to $\x_s$ in the discrete picture).

The implicit assumption in the above derivation of the Robin-type
boundary condition is that the transition rate $W_{b,s}$ to the bound
state scales as $1/a$ in the limit $a\to 0$, in order to get a finite
reactivity $\kappa_0$ in Eq. (\ref{eq:kappa_W}).  A more general
scaling
\begin{equation}  \label{eq:W_scaling}
W_{b,s} \sim a^{-\gamma}  \quad (a\to 0) 
\end{equation}
with $\gamma > 1$ would yield an infinite reactivity and thus the
Dirichlet boundary condition $c(\x,t) = 0$ at $\x\in\pa_R$.  Once the
particle hits such a point, it is immediately adsorbed.  Moreover, as
the desorption occurs at the same boundary point, the unbound particle
re-binds immediately.  In other words, this scaling leads to a perfect
irreversible binding, regardless the unbinding kinetics.  In turn,
scaling (\ref{eq:W_scaling}) with $\gamma < 1$ yields $\kappa_0 = 0$,
i.e., an inert reflecting boundary.  As the particle started from the
unbound state, one has $c_b(\x,0) = 0$ and thus the diffusive flux
remains zero at all times.  We stress that the scaling relation
(\ref{eq:kappa_W}) is indeed an assumption, in the same way as the
scaling relation (\ref{eq:D_a}) for the diffusion coefficient: if
Eq. (\ref{eq:D_a}) does not hold, the master equation does not
converge to the macroscopic diffusion equation; similarly, if
Eq. (\ref{eq:kappa_W}) is not valid, the macroscopic limit does not
represent a partially reactive boundary.

\section{Distribution of binding times}
\label{sec:binding_time}

The probability flux density $j_{\psi}(\x,t|\x_0)$ can be interpreted
as the joint probability density of the binding position $\X_{\T}$ and
the binding time $\T$ defined by Eq. (\ref{eq:Tdef}).  In particular,
the integral of this quantity over $\x\in \pa_R$ determines the
(marginal) probability density of the binding time
\begin{equation}
J_\psi(t|\x_0) = \int\limits_{\pa_R} d\x \, j_\psi(\x,t|\x_0).
\end{equation}
As a consequence, the spectral expansion (\ref{eq:jpsi}) implies
\begin{align}  \nonumber
\tilde{J}_\psi(p|\x_0) & = \E_{\x_0}\{ e^{-p\T}\} \\  \label{eq:Jpsi}
& = \sum\limits_{k=0}^\infty [V_k^{(p)}(\x_0)]^* \dtilde{\psi}(\mu_k^{(p)})  \int\limits_{\pa_R} d\x \, v_k^{(p)}(\x)\,.
\end{align}
This quantity determines all positive integer-order moments of the
binding time $\T$ (if they exist),
\begin{equation}
\E_{\x_0} \{ \T^n \} = (-1)^n \lim\limits_{p\to 0} \frac{\partial^n}{\partial p^n} \tilde{J}_\psi(p|\x_0) ,
\end{equation}
while its inverse Laplace transform yields the probability density
$J(t|\x_0)$.

Let us examine the small-$p$ asymptotic behavior of
Eq. (\ref{eq:Jpsi}) by treating separately the contributions from the
ground eigenmode $k = 0$ and other eigenmodes with $k > 0$.  In fact,
as $v_0^{(p)}$ approaches a constant in the limit $p\to 0$, the
integral in Eq. (\ref{eq:Jpsi}) vanishes for any $k > 0$.
Substituting the small-$p$ expansion (\ref{eq:psitilde_0}) for $k =
0$, we get
\begin{equation}  \label{eq:Jpsi_asympt0}
\tilde{J}_\psi(p|\x_0) \approx 1 - p^\alpha \ttau^\alpha + O(p)  \quad (p\to 0).
\end{equation}
If $0< \alpha < 1$, the correction $O(p)$ from all other terms is
subleading, and the mean binding time is infinite.  In this case,
binding events are characterized by the time scale $\ttau$ from
Eq. (\ref{eq:tau}).  In contrast, if $\alpha = 1$, both contributions
in Eq. (\ref{eq:Jpsi_asympt0}) are comparable and sum up to determine
the mean binding time.

\section{Diffusion on an interval}
\label{sec:interval}

In this Appendix, we summarize the formulas in the case of diffusion
on an interval $(0,L)$, with the reactive endpoint $x = 0$ and the
reflecting endpoint $x = L$ \cite{Grebenkov20b}.  In this case,
Eqs. (\ref{eq:g0}, \ref{eq:mu0_sphere}) are replaced by
\begin{align}
g_0^{(p)}(x) & = \frac{\cosh((L-x)\sqrt{p/D})}{\cosh(L\sqrt{p/D})} \,,\\  \label{eq:mu0_1D}
\mu_0^{(p)} & = \sqrt{p/D} \, \tanh(L\sqrt{p/D}) , 
\end{align}
while other relations in the beginning of Sec. \ref{sec:sphere} remain
unchanged, except for the diffusive flux, which reads
\begin{equation}
\tilde{J}(p) = (-D \partial_n \tilde{c}(x,p))_{x=0} 
= c_0 D \frac{\mu_0^{(p)} (1-\tilde{\phi}(p)) \dtilde{\psi}(\mu_0^{(p)})}
{p[1 - \tilde{\phi}(p) \dtilde{\psi}(\mu_0^{(p)})]} 
\end{equation} 
and differs from Eq. (\ref{eq:Jtilde_sphere}) by the factor $4\pi R^2$
(the surface area of the reactive region).  The major difference
between these two settings araises in the limit $L\to \infty$ when the
outer reflecting boundary is moved to infinity so that $\mu_{0}^{(p)}
= \sqrt{p/D}$ on the half-line.  As $p\to 0$, this eigenvalue
vanishes, whereas the eigenvalue $\mu_0^{(p)}$ from
Eq. (\ref{eq:mu0_sphere}) approached a strictly positive constant
$1/R$ (see Sec. \ref{sec:sphere}).  This distinction is related to the
transient (resp.  recurrent) character of three-dimensional
(resp. one-dimensional) Brownian motions, see
\cite{Grebenkov21} for further discussions.

\section{Short-time behavior}
\label{sec:Sshort_sphere}

We briefly discuss the short-time asymptotic behavior of $S(t|\x_0)$
for the case of diffusion between concentric spheres.  According to
Eq. (\ref{eq:g0}, \ref{eq:mu0_sphere}), one has $g_0^{(p)}(r_0)
\approx e^{-(r_0-R)\sqrt{p/D}} R/r_0$ and $\mu_0^{(p)} \approx
\sqrt{p/D}$ in the limit $p\to\infty$.  For the Mittag-Leffler model,
one gets
\begin{equation}  \label{eq:Upsilon_ML_large}
\dtilde{\psi}(\mu) \approx (\ell_0 \mu)^{-\alpha} \quad (\mu \to \infty) ,
\end{equation}
and so that Eq. (\ref{eq:Stilde_sphere}) implies
\begin{equation} 
\tilde{S}(p|\x_0) \approx \frac{1}{p} - e^{-(r_0-R)\sqrt{p/D}} \frac{R D^{\alpha/2}}{r_0 \ell_0^\alpha p^{1+\alpha/2}}
\quad (p\to\infty).
\end{equation}
Expectedly, this behavior does not depend on the waiting time
distribution.  In fact, the particle does not have enough time to
unbind in this limit, and the probability $S(t|\x_0)$ at short times
is mainly affected by the first binding event.

The inverse Laplace transform yields
\begin{align*}
S(t|\x_0) & \approx 1 - \frac{R D^{\alpha/2}}{r_0 \ell_0^\alpha} \int\limits_0^t dt_1 \, 
\frac{(r_0-R) e^{-(r_0-R)^2/(4Dt_1)}}{\sqrt{4\pi Dt_1^3}} \\
& \times \frac{(t-t_1)^{\alpha/2}}{\Gamma(\alpha/2+1)}  \quad (t\to 0).
\end{align*}
After simplifications, we get 
\begin{equation}
S(t|\x_0) \approx 1 - \frac{2^{1+\alpha} R \, e^{-(r_0-R)^2/(4Dt)} }{\sqrt{\pi}\, r_0 \,\ell_0^\alpha \,(r_0-R)^{\alpha+1}} 
\, (Dt)^{\alpha+1/2}  \quad (t\to 0).
\end{equation}
One sees that the short-time behavior is controlled by the first
arrival onto the reactive region (the typical exponential factor
$e^{-(r_0-R)^2/(4Dt)}$ from the L\'evy-Smirnov probability density of
the first-passage time) and by small random thresholds, i.e., by the
behavior of $\psi(\ell)$ as $\ell\to 0$.  The chosen Mittag-Leffler
distribution is responsible for the corrective power-law factor
$t^{\alpha+1/2}$.  We stress that if $\dtilde{\psi}(\mu)$ does not
exhibit a power-law decay (\ref{eq:Upsilon_ML_large}), the result will
be different.

\end{document}